\documentclass[twocolumn]{aastex631}

\usepackage{amsmath}
\usepackage{booktabs, threeparttable}
\usepackage{tabularx}

\newcommand{\dd}{\mathrm{d}}
\newcommand{\frbzero}{FRB20211127I}
\newcommand{\frbone}{FRB20211212A}
\newcommand{\frbtwo}{FRB20190608B}
\newcommand{\frbthree}{FRB20200430A}
\newcommand{\frbfour}{FRB20191001A}
\newcommand{\frbfive}{FRB20190714A}
\newcommand{\frbsix}{FRB20180924B}
\newcommand{\frbseven}{FRB20200906A}
\newcommand{\frbeight}{FRB20210117A}
\newcommand{\frbnine}{FRB20190520B}

\newcommand{\zfrb}{\ensuremath{z_{\rm FRB}}}

\newcommand{\dmunits}{\ensuremath{\rm pc \, cm^{-3}}}

\newcommand{\dmhalos}{\ensuremath{\mathrm{DM}_\mathrm{halos}}}

\newcommand{\dmfrb}{\ensuremath{\mathrm{DM}_\mathrm{FRB}}}
\newcommand{\dmigm}{\ensuremath{\mathrm{DM}_\mathrm{IGM}}}

\newcommand{\dmhost}{\ensuremath{\mathrm{DM}_\mathrm{host}}}

\newcommand{\dmmw}{\ensuremath{{\rm DM}_{\rm MW}}}

\newcommand{\dmcosmic}{\ensuremath{\mathrm{DM}_\mathrm{cosmic}}}

\newcommand{\mhalo}{\ensuremath{M_\mathrm{halo}}}

\newcommand{\figm}{\ensuremath{f_\mathrm{igm}}}

\newcommand{\wide}{\textsc{Wide}}
\newcommand{\narrow}{\textsc{Narrow}}
\newcommand{\ifu}{\textsc{Ifu}}

\begin{document}

\title{FRB Line-of-sight Ionization Measurement From Lightcone AAOmega Mapping Survey: \\the First Data Release}

\author[0000-0002-0298-8898]{Yuxin Huang}
\email{mochafhxy@gmail.com}
\affiliation{Kavli IPMU (WPI), UTIAS, The University of Tokyo, Kashiwa, Chiba 277-8583, Japan}

\author[0000-0003-3801-1496]{Sunil Simha}
\affiliation{University of California, Santa Cruz, 1156 High St., Santa Cruz, CA 95064, USA}

\author[0000-0003-0574-7421]{Ilya S. Khrykin}
\affiliation{Instituto de Física, Pontificia Universidad Católica de Valparaíso, Casilla 4059, Valparaíso, Chile}

\author[0000-0001-9299-5719]{Khee-Gan Lee}
\affiliation{Kavli IPMU (WPI), UTIAS, The University of Tokyo, Kashiwa, Chiba 277-8583, Japan}
\affiliation{Center for Data-Driven Discovery, Kavli IPMU (WPI), UTIAS, The University of Tokyo, Kashiwa, Chiba 277-8583, Japan}

\author[0000-0002-7738-6875]{J. Xavier Prochaska}
\affiliation{University of California, Santa Cruz, 1156 High St., Santa Cruz, CA 95064, USA}
\affiliation{Kavli IPMU (WPI), UTIAS, The University of Tokyo, Kashiwa, Chiba 277-8583, Japan}
\affiliation{Division of Science, National Astronomical Observatory of Japan, 2-21-1 Osawa, Mitaka, Tokyo 181-8588, Japan}
\affiliation{Simons Pivot Fellow}

\author[0000-0002-1883-4252]{Nicolas Tejos}
\affiliation{Instituto de Física, Pontificia Universidad Católica de Valparaíso, Casilla 4059, Valparaíso, Chile}

\author[0000-0003-2149-0363]{Keith W. Bannister}
\affiliation{Australia Telescope National Facility, CSIRO, Space and Astronomy, PO Box 76, Epping, NSW 1710, Australia}
\affiliation{Sydney Institute for Astronomy, School of Physics, The University of Sydney, NSW 2006, Australia}

\author{Jason Barrios}
\affiliation{University of California, Santa Barbara, Santa Barbara, CA 93106, USA}

\author[0000-0002-0302-2577]{John Chisholm}
\affiliation{Department of Astronomy, The University of Texas at Austin, 2515 Speedway, Stop C1400, Austin, TX 78712, USA}

\author[0000-0001-5703-2108]{Jeff Cooke}
\affiliation{Centre for Astrophysics and Supercomputing, Swinburne University of Technology, Hawthorn, VIC 3122, Australia}
\affiliation{ARC Centre of Excellence for All Sky Astrophysics in 3 Dimensions (ASTRO 3D), Australia}

\author[0000-0001-9434-3837]{Adam~T.~Deller}
\affiliation{Swinburne University of Technology}

\author[0000-0002-5067-8894]{Marcin Glowacki}
\affiliation{International Centre for Radio Astronomy Research (ICRAR), Curtin University, Bentley, WA 6102, Australia}

\author[0000-0003-1483-0147]{Lachlan Marnoch}
\affiliation{School of Mathematical and Physical Sciences, Macquarie University, NSW 2109, Australia}
\affiliation{Astrophysics and Space Technologies Research Centre, Macquarie University, Sydney, NSW 2109, Australia}
\affiliation{Australia Telescope National Facility, CSIRO Space \& Astronomy, Box 76 Epping, NSW 1710, Australia}
\affiliation{The ARC Centre of Excellence for All-Sky Astrophysics in 3 Dimensions (ASTRO 3D)}

\author[0000-0002-7285-6348]{R.~M.~Shannon}
\affiliation{Centre for Astrophysics and Supercomputing, Swinburne University of Technology, Hawthorn, VIC 3122, Australia}

\author[0000-0001-5310-4186]{Jielai Zhang}
\affiliation{Centre for Astrophysics and Supercomputing, Swinburne University of Technology, Hawthorn, VIC 3122, Australia}
\affiliation{ARC Centre of Excellence for Gravitational Wave Discovery (OzGrav), Hawthorn, VIC 3122, Australia}



\begin{abstract}
This paper presents the first public data release (DR1) of the FRB Line-of-sight Ionization Measurement From Lightcone AAOmega Mapping (FLIMFLAM) Survey, a wide field spectroscopic survey targeted 
on the fields of 10 precisely localized Fast Radio Bursts (FRBs). DR1 encompasses spectroscopic data for 10,468 galaxy redshifts across 10 FRBs fields with $z<0.4$, covering approximately $26\text{ deg}^2$ of the sky in total. 
FLIMFLAM is composed of several layers, encompassing the `Wide' (covering $\sim$ degree or $>10$ Mpc scales), `Narrow', (several-arcminute or $\sim$Mpc) and integral field unit (`IFU'; $\sim$ arcminute or $\sim 100$kpc ) components. The bulk of the data comprise spectroscopy from the 2dF-AAOmega on the 3.9-meter Anglo-Australian Telescope, while most of the Narrow and IFU data was achieved using an ensemble of 8-10-meter class telescopes. We summarize the information on our selected FRB fields, the criteria for target selection, methodologies employed for data reduction, spectral analysis processes, and an overview of our data products. An evaluation of our data reveals an average spectroscopic completeness of 48.43\%, with over 80\% of the observed targets having secure redshifts.
Additionally, we describe our approach on generating angular masks and calculating the target selection functions, setting the stage for the impending reconstruction of the matter density field.

\end{abstract}




\section{Introduction} \label{sec:intro}

Fast radio bursts (FRBs) are transient radio pulses that typically last for milliseconds, that have conclusively been shown to be extragalactic sources. A few low-luminosity examples are found to originate from our Galaxy but their relationship to the extragalactic population is still debated \citep{Andersen2020, Bochenek2020}. The extreme energies of FRBs imply they are generated from high-energy astrophysical processes that have not been fully understood yet, but many observations and theories are now converging on magnetars as the most likely engine for at least some of the FRB population \citep{Bochenek2020, CHIME+Andersen2020, Kirsten2021, Andersen2022, Kramer2024}. 

The first samples of FRBs were detected by large single-dish radio telescopes such as Parkes, Arecibo or the Green Bank Telescope, such that FRB positions could only be measured to within few arcminutes, which rendered it nearly impossible to localize a single FRB to a host galaxy.
Only with repeated bursts and interferometry
(e.g. FRB20121002, \citealt{Spitler2016}), did it become feasible to beat down the uncertainties of a given FRB's position to plausibly associate a galaxy as its host \citep{Eftekhari2017}.

Localized FRBs with known host galaxies were therefore a comparative rarity in the early years of FRB research.
However, wide-field interferometric radio arrays like the Australian Square Kilometre Array Pathfinder  \citep[ASKAP;][]{mcconnell_2016, Shannon2018TheDR} and the Deep Synoptic Array \citep[DSA;][]{10.1093/mnras/stz2219} have been systematically searching for FRBs over the past few years, with the position of detected FRBs constrained to within an arcsecond or better with just a single burst detection. This is sufficient to isolate them to individual host galaxies within the error circle of localization \citep{bannister2019}.
Since non-repeating FRBs dominate the overall observed sample, interferometric FRB surveys have significantly increased the number of localized FRBs, from just a handful in 2018 to nearly 100 at the time of writing, with approximately 45 published as of August 2024. 

The radio waves emitted by a FRB experience multiple propagation effects caused by the plasma along its path to Earth, like dispersion, scattering, scintillation, plasma lensing, absorption and Faraday rotation. Here we briefly introduce the dispersion effect, in which radio waves are dispersed by free electrons in the intervening plasma. Thus the arrival time, $\Delta t$ is different between photons with different frequencies $\nu_1$ and $\nu_2$, where

\begin{equation}
    \begin{split}
        \Delta t&=\frac{e^2}{2\pi m_e c}\left(\frac{1}{\nu_1^2}-\frac{1}{\nu_2^2}\right)\int\frac{n_e}{1+z}\dd l\\
                &\simeq 4.15\text{ s}\left[\left(\frac{\nu_1}{\text{GHz}} \right)^{-2}-\left(\frac{\nu_2}{\text{GHz}} \right)^{-2}\right]\frac{\text{DM}}{10^3\text{ pc cm}^{-3}},
    \end{split}
\end{equation}
where the dispersion measure 
\begin{equation}
    \mathrm{DM} \equiv\int \frac{n_e}{ 1+z}\,\dd l
\end{equation} is the integral over free electrons, $n_e$ along the line of sight, $\dd l$ is the proper line element, and $z$ is the corresponding redshift \citep{Lorimer2004}.
In a flat $\Lambda$CDM spacetime, the total DM generated by a line-of-sight distribution of $n_e$ by an FRB at redshift $z_{\text{frb}}$ is given by:
\begin{equation}
    \text{DM}=\frac{c}{H_0}\int_0^{z_{\text{frb}}}\frac{\dd z\text{ }n_e(z)}{(1+z)^2\sqrt{\Omega_m(1+z)^3+1-\Omega_m}},
\end{equation}
 $\Omega_m$ is the matter density of today's universe in units of $\rho_{c,0}=3H_0^2/8\pi G$ and $\rho_{c,0}$ is the critical density, and $c$ is the speed of light.

 The observed DM in extragalactic FRBs can typically be divided into four parts:

\begin{equation}\label{DM_components}
    \dmfrb=\dmmw+\dmigm+\dmhalos+\dmhost,
\end{equation}
where \dmmw{} is the contribution from the Milky Way, \dmigm{} is the dispersion measure of the diffuse IGM gas tracing the large-scale structure on $\geq$Mpc scales, \dmhalos{} comes from the dispersion from intervening galaxy halos within tens to a few hundreds of transverse kpc from the line of sight, and \dmhost{} is the contribution of the FRB host galaxy and source. While some authors use the term \dmigm{} to collectively refer to all IGM and CGM contributions beyond the Milky Way and host galaxy, we explicitly separate out these two contributions as \dmigm{} and \dmhalos{}; as a shorthand we define $\dmcosmic \equiv \dmigm + \dmhalos$. 

The Milky Way component is, in principle, well known based on models of the Galactic electron density distribution as probed by Galactic pulsars, although those models have systematic uncertainties due to HII regions that are not well-modeled. $\text{DM}_{\text{halos}}$ and $\text{DM}_{\text{host}}$ do not have specific predicted values from theory. 
In the former, this is largely due to our ignorance on the CGM gas fractions \citep{Sorini2022, Ayromlou2023, Khrykin2024simba} even if the foreground galaxies are well characterized. 
$\text{DM}_{\text{host}}$, on the other hand, has unknown contribution from the FRB engine as well as the host galaxy ISM and CGM in the initial portion of its path.

For FRBs at distances beyond the Local Universe ($\zfrb\gtrsim 0.1$ or several hundred Mpc), the DM budget of Equation~\ref{DM_components} is typically dominated by \dmhalos{} and/or \dmigm. 
\dmhalos{} is the contribution from the circumgalactic media (CGM) of intervening galaxies, typically within their virial or characteristic halo radii and residing in matter overdensities of $\rho/\langle \rho \rangle >100$.
Meanwhile, \dmigm{} is the contribution from the average-to-low density gas $0\lesssim \rho/\langle \rho \rangle \lesssim 10$ permeating the voids, sheets and filaments of the intergalactic cosmic web, which dominates the baryon budget.

The DM of FRBs have thus become a useful probe for studying both the extreme environments around the sources of FRBs, and the ionized gas along their propagating paths. Numerous papers have discussed using FRBs to quantify the large-scale structure theoretically \citep{PhysRevLett.115.121301}. With DMs of hypothetical samples of $\sim10^4$ FRBs it is possible to yield a significant clustering signal under the condition that local DM contributions are small \citep{Zhou_2014}. The idea of using large FRB samples to probe cosmological magnetic fields has also been suggested by \citet{10.1093/mnras/stx2830}. A more challenging idea is to use the DM-$z$ relation to measure the proper distance of the universe \citep{Yu_2017}, the equation of state of dark energy \citep{Walters_2018}, the Hubble constant \citep{Hagstotz_2022, Wu_2022}, the Hubble parameter \citep{Wu_2020}, the dark matter \citep{Wang_2018, Munoz2016} and the cosmic helium and hydrogen reionization history \citep{Zheng_2014}. These require large samples of localized FRBs with measured distances and for the different DM contributions to be separated well (at least statistically).  

One long-standing question that is particularly addressable with FRBs is that of the so-called missing baryon problem \citep{Fukugita_2004, Cen_2006, Bregman_2007}.
Within the framework of $\Lambda$CDM cosmology, methods such as Big Bang nucleosynthesis (BBN) and cosmic microwave background (CMB) temperature anisotropies have yielded tight constraints on the cosmological baryon fraction $\Omega_b$ \citep{Planck_2020}. However, it was pointed out \citep{Fukugita+98,Fukugita_2004}  that visible stars and ISM gas within observable galaxies constitute only $\sim 10\%$ of the cosmic baryon budget in the local universe ($z<0.5$).

The baryons in the CGM and IGM at $z\gtrsim2$ have been accounted-for in the observed population of HI Ly$\alpha$ absorbers due to the relatively simple physics of photoionization equilibrium that prevails in the quasi-linear structure formation regime of early times. The IGM at $z<1$, on the other hand, is predicted to be 
composed of a complex multi-phase medium due to the gravitational shock-heating and feedback from galaxy formation \citep{Cen_2006, Smith_2011}. Since baryons residing in this region can not be detected easily by X-ray emission or absorption-line tracers
\citep[e.g.][]{pt2009}, 
there were various heroic efforts to develop new observational techniques to solve this problem. Examples include observing HI Ly$\alpha$ absorption using the \textit{Hubble Space Telescope} \citep{Danforth_2008, Tejos2016}, utilizing high-ionization metal absorption \citep{Prochaska_2011, Nicastro_2018}, analyzing stacked X-ray emission in filaments \citep{Tanimura_2020} or using the Sunyaev-Zel'dovich effect \citep{Tanimura_2018, Tanimura_2019, WuXuanyi_2020}. Nevertheless, as of early 2020 $\approx 20\%$ of the cosmic baryons still remained undetected \citep{deGraaff2019}.

Since the vast majority of baryons residing in the CGM and IGM are fully ionized \citep{gp1965}, the cosmic free electron distribution is a simple and well-understood proxy for the baryon distribution in the local Universe, thus making FRBs the premier new technique for constraining the missing baryon problem.
 Using 7 FRBs that have been localized through the CRAFT/ASKAP survey, \citet{Macquart_2020} presented an extragalactic DM-$z$ relation (the so-called ``Macquart relation'') which is shown to be consistent with the value of $\Omega_b$ expected from $\Lambda$CDM cosmology. 
 This striking result showed that the hitherto undetected baryons expected from CMB and BBN constraints indeed reside within the ionized IGM and CGM gas intersected by the observed FRBs.

While the cosmic baryons are no longer ``missing'' from the Local Universe, their relative distribution within the Universe remains a very compelling question for both astrophysics and cosmology.
This question was initially formulated in terms of the large scatter of DM$_{\rm cosmic}$ around the mean value at a given redshift \citep{McQuinn2014,macquart2020}, which arises from the sightline-to-sightline variation in the intersected galactic halos as well under- and over-densities of the cosmic web along the line-of-sight \citep{Walker2024}.
The exact shape of DM$_{\rm cosmic}$ distribution at fixed redshift is believed to be partially shaped by the redistribution of cosmic baryons due to structure formation and galactic feedback (AGN, star-formation) \citep{Batten2022,Baptista2024,Medlock2024}.
Indeed, in a recent paper, \citet{Khrykin2024simba} showed using cosmological hydrodynamical simulations, that the primary large-scale effect of different feedback models is to eject gas (i.e.\  baryons) from galaxy halos into the IGM, with the retained halo gas fractions varying as a function of halo mass and feedback model (see also \citealt{Sorini2022, Ayromlou2023}). 
They showed measuring the gas fractions of different halo mass ranges as well as in the diffuse IGM could in principle be a powerful probe of feedback mechanisms.
A follow-up study by Dedieu et al (2024, in prep.) finds however that once gas is ejected out of halos, there is little further redistribution between the nodes, filaments, sheets and voids of the cosmic web.  

The DM$_{\rm cosmic}$ distribution as measured by FRBs is thus a promising probe of galaxy and AGN feedback, which remains one of the major questions in contemporary astrophysics, as well as potentially a source of systematic error for cosmological analyses. 
However, \citet{Batten2022} showed that samples of $\sim 10^3$ localized FRBs would be required to discern the effect of feedback on the \dmcosmic{} distribution. 
In a recent study, \citet{Baptista2024} modeled the variance in extragalactic dispersion measure using a fluctuation parameter. They projected that analyzing 100 localized FRBs would adequately constrain the fluctuation parameter's upper and lower limits to a $3\sigma$ level and determine $H_0$ to within 10\%, assuming a Gaussian prior on the logarithm of the fluctuation parameter.
With \textit{unlocalized} FRBs that have $\sim$arcminute positional uncertainties and unknown FRB redshift, such constraints are even more challenging: \citet{Shirasaki2022} estimated that $\sim 20,000$ unlocalized FRBs would be required to constrain halo gas mass fractions in cross-correlation with foreground spectroscopic samples (see also \citealt{Wu2023}).

In \citet{Lee_2022}, we proposed to combine localized FRBs with detailed spectroscopic information on potential galactic and cosmic contributions to their DM, which we dubbed ``FRB foreground mapping". 
While it was clear early on that observations of intervening galactic halos in front FRBs could yield interesting information on the halo DM contributions \citep{Prochaska2019_FRB181112,Simha2021}, \citet{Lee_2022} show that wider spectroscopic surveys across degree-scales on the sky would allow the large-scale matter density field to be reconstructed. 
This would especially allow much tighter constraints on the \dmigm{} component as shown in Eq.\ref{DM_components}. 
To a good approximation (see \citealt{Lee_2022}),  the IGM free electron density, $n_{e,\text{igm}}$, can be related to the underlying total matter density, $n_{\text{lss}}$, through 
\begin{equation}
n_{e,\text{igm}}(l)=f_{\text{igm}}A_{\text{bar}}\,n_{\text{lss}}(l)
\end{equation}
where $A_{\text{bar}}$ takes into account the electrons contributed by the primordial hydrogen and helium abundances, $f_{\text{igm}}$ is a free parameter governing the fraction of cosmic baryons located in IGM. 
Given the foreground distribution of galaxies along the line of sight $l$, we are able to reconstruct the underlying matter density field of the large-scale structure and compute models of the \dmigm{} contribution as a function of $f_{\text{igm}}$. 
\citet{Lee_2022} argued that the direct modeling of per-sightline \dmigm{}, alongside the other terms in Equation~\ref{DM_components}, dramatically increases the constraining power of localized FRBs towards \figm{} as well as the halo gas fractions.

\citet{Simha2020} has carried out a preliminary implementation of this idea. Using spectroscopic data from the Sloan Digital Sky Survey (SDSS) in the foreground of FRB20190608B ($\zfrb=0.1167$), they estimated the contribution of hot ionized gas in the intervening halos and derived $\dmhalos\approx 7-28$ pc cm$^{-3}$. They then used the Monte Carlo Physarum Machine (MCPM) algorithm \citep{Burchett2020} to produce a 3D map of ionized gas in cosmic web filaments and calculated the IGM contribution of $\dmigm\approx 91-126$ pc cm$^{-3}$. They concluded that there are a greater fraction of ionized gas extant outside virialized halos along the line of sight of FRBs. Although this analysis was limited by the small sample of one FRB, they confirmed the feasibility of this method to constrain \dmhalos{} and \dmigm{} and subsequently constrain the cosmic baryon distribution. \cite{Lee_2022} showed that with the increasing sample of localized FRBs over the next few years, this approach can make precise constraints on the cosmic baryon distribution.

To obtain spectroscopic observational data to implement this idea across a sample of localized FRBs, we initiated the FRB Line-of-sight Ionization Measurement From Lightcone AAOmega Mapping (FLIMFLAM) Survey. As outlined by \citet{Lee_2022}, our approach employs a tripartite observing strategy designed to constrain the free parameters in the extragalactic DM components: \dmhalos{} and \dmigm. 
To map out the cosmic web across transverse scales of $\gtrsim 10$Mpc in the FRB forground, we require massively-multiplexed galaxy spectroscopic data
targeting relatively bright galaxies to act as large-scale structure tracers. 
For a subset of our localized FRB sample, such data was already publicly available from surveys like the Sloan Digital Sky Survey (SDSS, \citealt{Abazajian2009}) or 6dF \citep{6dF2009}. However, many of our FRBs are in the Southern Hemisphere, where SDSS does not cover, or at a redshift ($\zfrb \gtrsim 0.1$) beyond that covered by 6dF. 
We therefore conducted a multi-year campaign on the 3.9m Anglo-Australian Telescope (AAT) to obtain the requisite wide-field spectroscopic data in front of a sample of FRBs' hosts. 
Simultaneously, we also secured observing time on several 8-10m class  telescopes (Keck-I/II, Gemini North) to characterize fainter galaxies whose CGM haloes might directly intersect the FRB sightlines within several arcminutes.
Finally, we used 8-10m-class integral field unit (IFU) spectroscopy to target the faintest galaxies that might contribute to the FRB DMs inside $\sim 1$ arc-minute of the sightlines.

This paper documents the first data release (DR1) of the FLIMFLAM survey, which comprises 10 FRBs detected primarily by the CRAFT/ASKAP Survey and localized to their hosts by the F$^4$ collaboration \citep{shannon2024}. 
The primary scientific result from this data set was presented in a companion paper by \citet{Khrykin2024flimflam}, which provided the first-ever constraints on the global IGM and CGM baryon partition from FRBs. 
Due to the non-trivial observational strategy on multiple facilities and sizeable data sample (10984 spectroscopic redshifts in total), we provide this separate paper to describe the data set in detail. 
Other papers which used the data herein include: (i) \citet{Simha2023} who analyzed 4 FRB sightlines that exhibited elevated DM values given their redshift, and (ii) \citet{Lee_2023} who discovered that the huge DM excess ($\dmhost \approx 900\,\dmunits$, as claimed) in the well-known FRB20190520B is not primarily due to the host, but in fact caused at least partially by two foreground galaxy clusters. 
Here, we will provide detailed descriptions of the observations and resulting data used in the aforementioned papers, with the intention that the community should be able to reproduce these results.
The data will be publicly released on Zenodo after this paper completes the refereeing process.

\section{Survey Design} \label{sec:design}

\begin{figure}[htbp]
\centering
\includegraphics[width=0.45\textwidth]{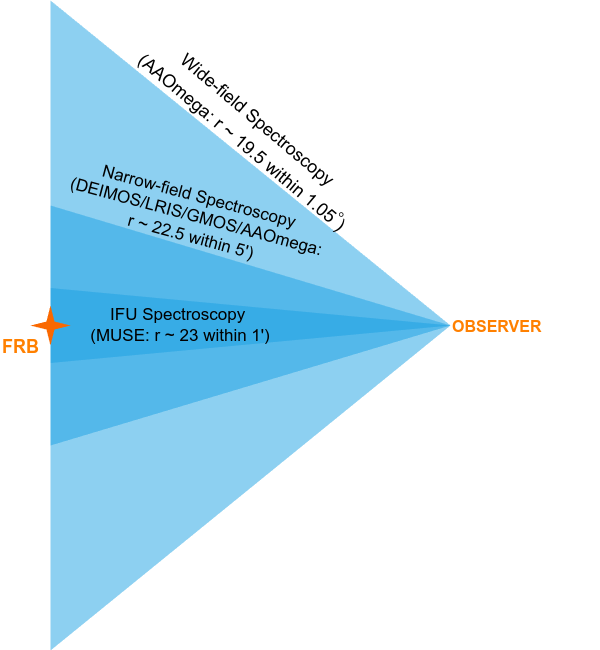}
\caption{A schematic illustration of the FLIMFLAM survey strategy (not to scale), which is generally organized into 3 survey layers. The \wide{} survey covers the widest FoV but is the shallowest and is observed using AAT as well as publicly-available SDSS and 6dF data. The \narrow{} survey, with moderate FoV and depth, is observed by Keck-I/II and Gemini. The \ifu{} component uses Keck/KCWI and VLT/MUSE to observe faintest targets closest to the sightline.}
\label{ff_survey}
\end{figure}

The FLIMFLAM survey is designed to map the foreground galaxies of localized FRBs, in order to constrain the baryon distribution in the local Universe ($z<0.5$) using density field reconstruction and parameter analysis techniques \citep{Lee_2022,Khrykin2024flimflam}. In this section, we describe the survey design, including the target selection criteria and the geometry of the observations on sky.

\subsection{FRB Sample} \label{subsec:sample}

All of the FRBs comprising this DR1 sample of 
FLIMFLAM were discovered by the CRAFT survey on the
ASKAP telescope \citep{James2019,shannon2024}.
That survey has now detected 63 FRBs and
has localized 37 to better than 1~arcsecond
precision.  Of these, we have restricted DR1 to 
FRBs with a host galaxy association posterior probability
$P(O|x) > 95\%$ using the PATH algorithm
\cite{Aggarwal_2021}.
Furthermore, we have avoided any FRB that lies too
close to the Galactic disk ($|b| < 30^\circ$)
or any that suffers from significant Galactic
extinction ($E(B-V) > 0.06$\,mag).
We also selected FRBs with redshifts
$0.04 < z_{\rm FRB} < 0.4$ to have significant
path-length through the cosmic web but still be
sensitive to a majority of the foreground galaxies
with luminosity $L > 0.1 L*$ (see following sections
for further details).
Last, we avoided several FRBs whose properties
suggested high local DM contributions.
Table~\ref{FRB_info} lists the sample including the
FRB dispersion measure $\rm DM_{FRB}$ which have
very small uncertainties (typically $<1 \dmunits$).

\subsection{Overall Strategy} \label{subsec:structure}

\begin{table*}[htbp] 
\small
\centering
\begin{threeparttable}
\begin{tabularx}{1.0\textwidth}{c|ccccccc}
\toprule
FRB & R.A. & Dec. & $z_{\rm FRB}$ & \wide{} & \narrow{} & \ifu{} data & Ref.  \\ 
    &(deg)&(deg)&          & data & data   &             &    \\ 
\midrule
20211127I  &  $199.8088$ & $-18.8381$ & $0.0469$ & 6dF   & AAT  &     /      &  \citet{2023ApJ...949...25G} \\
20211212A & $157.3507$ & $+01.3605$ & $0.0713$ & SDSS  & AAT  &     /      &  Deller et al. (in prep) \\
20190608B & $334.0199$ & $-07.8982$ & $0.1178$ & SDSS, 6dF & DEIMOS &   MUSE, KCWI   &  \citet{macquart2020} \\
20200430A & $229.7066$ & $+12.3761$ & $0.1608$ & SDSS, AAT & AAT, DEIMOS  &   MUSE   & \citet{heintz2020} \\
20191001A & $323.3516$ & $-54.7477$ & $0.2340$ & AAT & AAT, GMOS  &  MUSE     & \citet{heintz2020} \\  
20190714A & $183.9795$ & $-13.0207$ & $0.2365$ & AAT       & AAT, LRIS, DEIMOS  &  MUSE  & \citet{heintz2020}  \\
20180924B & $326.1052$ & $-40.9000$ & $0.3212$ & AAT       & AAT  &   MUSE     & \citet{bannister2019} \\
20200906A & $053.4956$ & $-14.0833$ & $0.3688$ & AAT       & AAT, DEIMOS  &   MUSE   & \citet{bhandari2020} \\
\midrule
20210117A & $339.9792$ & $-16.1515$ & $0.2145$ & AAT & DEIMOS & /  & \citet{Bhandari:2022ton}  \\
20190520B & $240.5178$ & $-11.2881$ & $0.2410$ & AAT & / & /  &  \citet{Niu2022} \\
\bottomrule
\end{tabularx}
\end{threeparttable}
\caption{A list that describes the basic information of the FLIMFLAM DR1 FRB fields. We list the Right Ascension (R.A.), Declination (Dec.) and redshfit of each FRB. We also show the surveys or instruments that we used to obtain the spectroscopic data for each branch survey in ``\wide{} data", ``\narrow{} data" and ``\ifu{} data" column. The FRB fields on the last two rows, separated by the horizontal line, are not included in the cosmic baryon analysis of \citet{Khrykin2024flimflam}.}
\label{FRB_info}
\end{table*}

The FLIMFLAM survey is composed of the wide-field (`\wide'), the narrow-field (`\narrow') and the integral field unit (`\ifu') survey layers, as illustrated in Fig.\ref{ff_survey}. 
This observational strategy is motivated by the decomposition of the FRB DM as described in Equation~\ref{DM_components}.

The \wide{} survey is comprised of relatively bright ($L \sim L^*$) galaxies that trace the large-scale cosmic web spanning tens of Megaparsecs, requiring observations of $\gtrsim 1000$ galaxies over at least several degrees on the sky. 
In conjunction with density reconstruction algorithms such as \texttt{ARGO} \citep{Ata_2015}, this allows us to recover the underlying 3D matter density and hence model \dmigm.
The core of FLIMFLAM is the \wide-AAT observational campaign targeting FRB foreground fields within the $1.05^\circ$ radius field-of-view (FoV) of the AAT/AAOmega fiber-fed spectrograph \citep{aaomega}. 
Four of the FRB fields are additionally covered by the Sloan Digital Sky Survey \citep[SDSS;][]{SDSS2023} and the 6-degree Field \citep[6dF;][]{6dF2009} spectroscopic surveys. Thus, we have augmented our \wide{} data with the publicly available redshifts from these surveys, which we refer to as \wide-SDSS and \wide-6dF respectively.

The \narrow{} and \ifu{} survey layers, on the other hand, are aimed at identifying and characterizing galaxies or haloes directly intersected by the FRB sightline, that might contribute to \dmhalos. 
In practice, galaxies detected by either of these survey components are treated similarly through the calculation of \dmhalos, but for operational clarity we maintain the distinction between \narrow{} and \ifu{} based on the nature of the instrumentation used.
In general, we aimed to execute the \narrow{} component using multi-object slitmasks spanning several arcminutes around each FRB on 8-10m class telescopes.
This included the Low-Resolution Imaging Spectrograph \citep[LRIS;][]{LRIS1995,LRIS2010,LRIS2022} and the DEep Imaging Multi-Object Spectrograph \citep[DEIMOS;][]{DEIMOS2003} on the 10.3m Keck Telescopes at the W. M. Keck Observatory for targets visible from the Northern Hemisphere. For southern fields, the Gemini Multi-Object Spectrograph \citep[GMOS;][]{GMOS2004} at the 8.2m Gemini South Telescope was used.
We denote these observations as `\narrow-8m'. 
However, some of our \narrow{} observations also utilized deeper exposures on the smaller AAT/AAOmega as a stopgap measure, as we will explain more detail below --- this data is referred to as `\narrow-AAT'.
Finally, the \ifu{} survey component was conducted through the Multi-Unit Spectroscopic Explorer \citep[MUSE;][]{MUSE2010} at the 8.2m Very Large Telescope (VLT) and the Keck Cosmic Web Imager \citep[KCWI;][]{KCWI2018} at Keck to target the faintest galaxies within $1'$ of the FRB sightlines. 

The first 8 FRB fields in Table \ref{FRB_info} are included in our cosmic baryon analysis \citep{Khrykin2024flimflam}. 
\frbzero{} and \frbone{} are at sufficiently low redshift ($\zfrb<0.1$) such that even relatively shallow ($\sim$19th mag) observations with the 3.9m AAT are sufficient to capture $L\sim 0.1\,L_*$ dwarf galaxies in their \narrow{} foreground, i.e. \narrow{} or \ifu{} observations with larger telescopes were deemed unnecessary. 
For each of these fields, we designed two AAOmega plates with a nominal depth of $r=19.8$ for galaxies within an 30$\arcmin$ radius of the FRB.
The \wide{} data for these two FRBs, on the other hand, were incorporated from the public SDSS and 6dF data sets.
\frbtwo{} also has a relatively low redshift ($\zfrb=0.1178$) and its \wide{} foreground is covered by both SDSS and 6dF, so we did not observe it with AAT at all; the \narrow{} and \ifu{} components for this FRB was observed using the DEIMOS and KCWI on the Keck-II telescope, and the MUSE on the VLT. 
\frbeight{} and \frbnine{} were `targets of opportunity' only analyzed as excess DM sources \citep{Simha2023,Lee_2023} and were not included in the detailed \citet{Khrykin2024flimflam} cosmic baryon analysis, so we did not observe them in the \narrow{} and \ifu{} components. These two FRBs are also at slightly lower Galactic latitudes and experienced greater Galactic extinction than the other fields, leading to lower effective spectral signal-to-noise and redshift success rate.

By design, all FLIMFLAM DR1 fields have \wide{} data covering multiple square degrees in the foreground, whether through our own AAT observations or from SDSS and 6dF. 
We have endeavored to also obtain 8-10m class \narrow{}-8m data within several arcminutes of the $\zfrb\gtrsim 0.1$ DR1 FRBs to be used for cosmic baryon analysis, but this is missing for \frbsix{} as of DR1. 
However, at the beginning of our survey, when we had AAT time in hand but no allocations on larger telescopes, as a stopgap we targeted fainter galaxies beyond the \wide{} thresholds within a $2.5\arcmin$ radius from the FRB. 
These fainter galaxies (nominally to a limit of $r=21.5$) were targeted with AAOmega across multiple plate configurations in order to go deeper than the \wide{} galaxies.
While these \narrow-AAT observations are shallower than the desired $r\sim 22-23$ depth for the \narrow{} galaxies, these were a useful stopgap.
The \narrow-AAT observations are the only \narrow{} data available for \frbsix{} as of DR1.
This means that we are possibly missing \dmhalos{} contributions from $L<0.3\,L_*$ galaxies in the \frbsix{} field, despite the central 1 arcsecond region being supplemented by MUSE data. However, considering the significant uncertainties in the DR1 cosmic baryon constraints \citep{Khrykin2024flimflam}, it is unlikely that this omission would qualitatively bias the DR1 results.

Finally, we have \ifu{} data on only 6 of our 8 DR1 fields analyzed by \citet{Khrykin2024flimflam}, but as mentioned before the \narrow{} data obtained with AAT on the two lowest-redshift FRBs (\frbzero{} and \frbone{}) functionally already reaches dwarf galaxy depths in their respective foregrounds.

\subsection{Target Selection}\label{subsec:selection}

Prior to the spectroscopic observations, our first step was to obtain imaging source catalogs for each FRB field in order to select targets. 
Due to the commensal nature of the CRAFT observations, the FRB positions are not contained within any single imaging survey footprint.
We make use of several publicly-available imaging surveys including the Dark Energy Survey (DES) DR1 \citep{Abbott_2018}, the Dark Energy Camera Legacy Surveys (DECaLS) DR8 \citep{2016AAS...22831701B}, the DECam Local Volume Exploration Survey (DELVE) DR2 \citep{Drlica-Wagne_2022}, and the Panoramic Survey Telescope and Rapid Response System (Pan-STARRS) DR1 \citep{Chambers_2016} survey. 
For \frbnine{}, no publicly available imaging was available prior to our spectroscopic observations, so we carried out our own imaging observations on the Blanco/DECam instrument to obtain the 
$r$-band magnitude of galaxies over a 2.2 degree field centered on the FRB, which is well-matched for follow-up spectroscopy with the 2 degree field on AAOmega. 


From each imaging catalog, we downloaded object sky coordinates (R.A., Dec.), Kron magnitudes and errors in the available bands: usually \emph{grizy} but also WISE (\textit{W1-W4}), SDSS \textit{u} and VISTA \textit{JHKs}. We used Kron magnitudes \citep{1980ApJS...43..305K} for all our analyses to ensure a uniform aperture photometry system was used across the various surveys. In addition, we obtained galaxy-star separation flags or SExtractor-based point source classification probability. In the case of DES data, we also obtained the photometric signal noise ratio (SNR) and excluded objects that had an $\rm SNR<5$ in $g,r,z$-band.

Our analysis relies on galaxies as tracers of the cosmic matter density fields. Thus, we excluded stars from our photometry tables. We then applied the following selection criteria: For DES catalogs, we directly used the SExtractor \citep{SExtractor} \texttt{class\_star} values of $g,r,z$-band and excluded objects with $\texttt{class\_star} > 0.5$. As for DECaLS catalogs, we used the morphological model fitted by \texttt{Tractor} \citep{tractor} and excluded the objects with \texttt{type} equal to ``PFS". In the case of DELVE objects, we used the spread model \citep{spreadmodel} star-galaxy classifier \texttt{extended\_class} of $g,r,i,z$-band. Finally for Pan-STARRS objects, we separate the stars from galaxies by calculating the difference between PSF and Kron magnitudes. We regard objects with \texttt{$i$PSFMag}$-$\texttt{$i$KronMag}$ < 0.05$ as stars where the \texttt{$i$PSFMag} is the $i$-band PSF magnitudes and \texttt{$i$KronMag} represents the $i$-band Kron magnitude \citep{2014MNRAS.437..748F}. After excluding stars, we de-reddened the photometry of galaxies using the \citet{1998ApJ...500..525S} Milky Way extinction map.

\begin{table*}[htbp] 
\small
\centering
\begin{threeparttable}
\begin{tabularx}{1.\textwidth}{c|cccccccc}
\toprule
FRB  & Source & \wide-AAT & \wide-AAT & \wide-AAT & \narrow-AAT & \narrow-AAT & \narrow-AAT \\ 
 & & Field Radius & $r_{\text{cut}}$ & $r_{\text{eff}}$ & Field Radius & $r_{\text{cut}}$ & $r_{\text{eff}}$ \\
\midrule
20211127I  & DELVE & / & / & / & $30'$ & 19.8 & 19.2  \\
20211212A  & DELVE & / & / & / & $30'$ & 19.8 & 19.2 \\
20200430A  & Pan-STARRS & $1.05^{\circ}$ & 19.2 & 18.6 & $2.5'$ & 21.5 & 21.5 \\
20191001A  & DES & $1.05^{\circ}$ & 19.4 & 19.4 & $2.5'$ & 21.5 & 21.5 \\  
20190714A  & Pan-STARRS & $1.05^{\circ}$ & 19.4 & 19.4 & $2.5'$ & 21.5 & 21.5 \\
20180924B  & DES & $1.05^{\circ}$ & 19.8 & 19.8 & $2.5'$ & 21.5 & 21.5 \\
20200906A  & DES & $1.05^{\circ}$ & 19.8 & 19.8 & $2.5'$ & 21.5 & 21.5 \\
\midrule
20210117A  & Pan-STARRS & $1.05^{\circ}$ & 19.4 & 19.4 & / & / & / \\
20190520B  & FF DECam & $1.05^{\circ}$ & 19.4 & 19.0 & / & / & / \\
\bottomrule
\end{tabularx}
\end{threeparttable}
\caption{This table details the imaging source and the radius of both the \wide-AAT and the \narrow-AAT fields. It also specifies the nominal upper $r$-band magnitude limit $r_{\text{cut}}$ used for \wide-AAT and \narrow-AAT target selection, as well as the effective depth $r_{\text{eff}}$ achieved in the \wide-AAT and \narrow-AAT surveys.}
\label{FRB_select}
\end{table*}

Subsequently, we impose an $r$-band magnitude cut on the apparent AB magnitudes as a simple way to select galaxies for spectroscopy. We implemented different magnitude cuts for FRBs at different redshifts \citep[see][for more details]{Lee_2022}, i.e. higher redshift FRBs requires greater depth to capture the bulk of their foreground galaxies. For FRBs localized at $\zfrb\lesssim[0.15,0.25,0.4]$, we used magnitude upper limits of $r\lesssim[19.2,19.4,19.8]$, respectively.
These magnitude thresholds were selected to be the same as those adopted by the Galaxy And Mass Assembly (GAMA) survey \citep{driver2011} which used the same AAOmega instrument as our program --- this will allow us to use the public GAMA survey data for cross-validation and estimation of radial selection functions of our own data.
For \frbthree, \frbfour, \frbfive, \frbsix{} and \frbseven, within a radius $2.5'$ circular field centered by the FRB, we apply a deeper magnitude cut $r\lesssim21.5$ to select \narrow-AAT targets for possible intervening galaxies. In Table \ref{FRB_select}, we list the nominal $r$-band magnitude upper limit $r_\text{cut}$ of each field in the \wide-AAT survey. 

During the early stages of the survey, we used circular aperture magnitudes with an aperture of $2"$ diameter from DES, DECaLS, and Pan-STARRS database to select targets. However, upon comparing the magnitudes of the same objects across different surveys within overlapping footprints, we found it more consistent to use Kron magnitudes as this definition takes into account the effective point-spread function (ePSF) of the telescopes and observing conditions across surveys. 
Here onwards, ``magnitudes'' refer to Kron magnitudes exclusively. Switching from aperture to Kron magnitude altered the effective depth of our target pool from the nominal magnitude threshold, and also excluded some previously observed galaxies. 
The effective depth $r_{\text{eff}}$ of each field is listed in Table \ref{FRB_select}. 
Generally, the effective depths were reduced by $\sim0.1$, thus excluding a small fraction of the observed galaxies ($\sim7\% $) from the target pool. 
In the case of \frbthree, however, the effective depth changed significantly to $r<18.6$. For \frbzero{ }and \frbone, the discrepancy between nominal and effective observed depth arises as we observed only the brighter $\sim40\%$ of our targets, limited by available observation time. However, given the very low redshifts, this brighter sample already reaches an approximate threshold $L\sim 0.1\, L^\star$ for possible foreground galaxies in these fields.

In the case of \frbnine{}, we carried out our own $r$-band source imaging with Blanco/DECam only $\sim 1$ month before the corresponding spectroscopic campaign. We were forced to select targets with preliminary image reductions \textit{sans} magnitude zero points.
For this field, we therefore defined a relative magnitude cut to select the 1500 brightest galaxies within a 2 degree field, which is the mean galaxy density expected for a nominal selection threshold of $r<19.4$ corresponding to galaxies in the foreground of a $\zfrb=0.25$ as is appropriate for the redshift of \frbnine\ ($\zfrb=0.241$).
When the zero-point was computed after the spectroscopic observations, the effective selection threshold turned out to be $r<19.1$ which after the (significant) dust correction corresponds to an unextincted magnitude of $r=18.4$ at the center of the field. As discussed in \citet{Lee_2023}, the large number of galaxies at this shallow threshold attests to multiple overdensities in the \frbnine{} field, including two galaxy clusters directly intersected by the FRB sightline.

As expected, the total number of selected \wide-AAT targets correlates positively with the $r$-band magnitude cut. The 4 FRB fields with a magnitude cut of 19.4 \frbthree, \frbfour, \frbeight{} and \frbnine{} have 9788 targets in total, which is 706 targets per deg$^2$ in average. The 2 fields with a magnitude cut of 19.8 \frbfive{} and \frbsix{} contain 7494 targets in total, namely 1082 targets per deg$^2$ in average. This target number is consistent with the GAMA survey, which has 669 targets per deg$^2$ in average for $r<19.4$ selections and 1047 targets per deg$^2$ for $r<19.8$ selections \citep{Baldry2010}.

As expected, the number of targets increases with the $r$-band magnitude cut, and the number of targets at each magnitude threshold are similar between the different fields. With this relation, we are able to estimate the rough number of targets with different magnitude cuts and predict the required observing time before we actually make the target selection. Furthermore, after we select the targets, we were able compare the target number with this curve to get a rough sense of the selection quality. For example, if the number of selected targets within a given field is much more than others with the same magnitude cut, it might be a hint that the selection (possibly the galaxy-star separation) is poor.

For the \narrow{}-AAT targets, given that a significant portion of galaxies in this magnitude interval have redshifts greater than \zfrb{} \citep{Lee_2022}, an additional selection criterion with photometric redshift (photo-$z$) was implemented to select objects located in the foreground.  We feed the Kron magnitudes and the corresponding error to \texttt{EAZY}  \citep{Brammer2008}, a photometric redshift estimation code\footnote{\href{https://github.com/gbrammer/eazy-photoz}{https://github.com/gbrammer/eazy-photoz}}, to obtain the photo-$z$ of the selected potential targets. Since only halos that lie in front of the FRBs can contribute to the the intervening halo contribution of DM, we only target for objects with photo-$z$, $z_{\rm phot}<\zfrb$. Considering the uncertainty of the photometric redshift estimation, we selected targets which satisfied $l95<\zfrb$, where $l95$ is the lower 95\% confidence bound in the photo-$z$. Specifically, we prioritized the observations of objects with photo-$z$'s satisfying $l68<\zfrb$ where $l68$ is the lower 68\% confidence bound.

For the \narrow-8m observations, we targeted galaxies within $\sim5$ arcmin of the fields of \frbtwo, \frbthree, \frbfour, \frbfive, \frbsix, \frbseven, and \frbeight. 
We use magnitude and color criteria to select our targets instead of photo-$z$.
For LRIS, DEIMOS, and GMOS observations, $r<23$ galaxies were selected from Pan-STARRS DR1 \citep{Chambers_2016}, where available, and with the NOIRLab Source Catalog (NSC) \citep{nsc_dr1,nsc_dr2} for southern fields outside the PS1 footprint. Based on both HectoMap \citep{Sohn2021} and VIPERS \citep{Guzzo2014} spectroscopic catalogs, we rejected objects which satisfied the following conservative color criteria to reduce the number of higher redshift ($z\gtrsim0.3$) sources:
\begin{equation}
    \begin{aligned}
        r-i & >0.7 \\
        i & >20.5
    \end{aligned}
\end{equation}

\begin{table*}[htbp] 
\centering
{\small
\begin{threeparttable}
\begin{tabularx}{0.88\textwidth}{c|cc|cc|cc|cc|cc}
\toprule
 & \multicolumn{2}{c|}{2020} & \multicolumn{2}{c|}{2021} & \multicolumn{2}{c|}{2022} & \multicolumn{2}{c|}{2023} & \multicolumn{2}{c}{Total} \\ 
FRB & plates & targets & plates & targets & plates & targets & plates & targets & plates & targets \\
\midrule
20211127I & 0 & 0 & 0 & 0 & 0 & 0 & 1 & 320 & 1 & (320) \\
20211212A & 0 & 0 & 0 & 0 & 0 & 0 & 1 & 285 & 1 & (285) \\
20200430A & 0 & 0 & 2 & 279(20) & 0 & 0 & 0 & 0 & 2 & 279(20) \\
20191001A & 9 & 2063(33) & 0 & 0 & 0 & 0 & 0 & 0 & 9 & 2063(33) \\
20190714A & 0 & 0 & 5 & 1451(14) & 2 & 458(5) & 0 & 0 & 7 & 1909(18) \\
20180924B & 8 & 2721(25) & 0 & 0 & 0 & 0 & 0 & 0 & 8 & 2721(25) \\
20200906A & 3 & 982(25) & 6 & 1750(15) & 0 & 0 & 0 & 0 & 9 & 2732(31) \\
20210117A & 0 & 0 & 5 & 1691 & 0 & 0 & 0 & 0 & 5 & 1691 \\
20190520B & 0 & 0 & 3 & 821 & 0 & 0 & 0 & 0 & 3 & 821 \\
\midrule
Total & 20 & 5766(83) & 21 & 5992(49) & 2 & 458(5) & 2 & (605) & 45 & 12216(732) \\
\bottomrule
\end{tabularx}
\end{threeparttable}
}
\caption{The AAT observing log for the FLIMFLAM DR1 survey. The table lists the amount of plates and both the \wide-AAT and the \narrow-AAT targets observed each year for each FRB field. The numbers outside the brackets represent the \wide-AAT targets, while the numbers within brackets indicate the \narrow-AAT targets. Note that for each FRB field, the total amount of \narrow-AAT targets may not equal the annual sums, as \narrow-AAT targets are observed up to three times, potentially across different observation runs.}
\label{logtab}
\end{table*}

\section{AAOmega observations and data reduction} \label{sec:obs}

The core of the FLIMFLAM survey is data from the AAOmega fiber spectrograph fed by the 2dF fiber positioner on the 3.9m AAT at Siding Spring Observatory, Australia. Our DR1 observations with the AAT covered 9 FRB fields, cumulatively spanning 25.82 deg$^2$ of the sky and a total of 12948 targets. Here, we describe our observation and data reduction methods.

\subsection{Observing with 2dF-AAOmega}\label{subsec:2df}

The 2dF-AAOmega system comprises a robotic fiber positioner and optical spectrograph that are often used together, but constitute two separate systems on the AAT. AAOmega is a dual-beam spectrograph, utilizing science fibers that converge into a single collimator. The light is then divided into blue and red spectral arms via a dichroic beam splitter \citep{2004SPIE.5492..410S}. For the FLIMFLAM survey, we chose to split the beam at 570nm. For the blue arm, we adopted the 580V grating and tune the central wavelength to be 4850\AA, while the red arm used the 385R grating with a central wavelength of 7250\AA. We set these values to ensure sufficient overlap between spectra from the red and blue arms during data reduction.

The 2dF fiber positioner is a multi-object system capable of acquiring up to 392 simultaneous spectra of objects across a two-degree diameter field on-sky \citep{2002MNRAS.333..279L}. It also has a wide-field corrector, an atmospheric dispersion corrector, and a robot gantry that can position fibers with an accuracy of $\sim0.3$ arcseconds on the sky. Two fiber plug plates are installed front and back on a tumbling mechanism at the telescope prime focus, which allows the next field to be configured during the current field is being observed.

For calibrations, we take 10 ``biases", an 8-second ``fiber flat" for the red arm, a 40-second ``fiber flat" for the blue arm, and a 45-second ``arc" frame. We used the Copper-Argon (CuAr), Iron-Argon (FeAr), Helium (He) and Neon (Ne) arc lamps. 

As for exposure times, for targets selected with magnitude cuts of $r=[19.2, 19.4, 19.8, 20.6]$ observed at airmass $\sim1.2$, we aim to take 3 exposures with nominal exposure times [700s, 800s, 900s, 1200s] respectively. For the \narrow-AAT targets, we assign fibers to three different plate configurations in order to triple the exposure times. We roughly scale the actual exposure times with the airmass during the observations, as well as adjustments based on observing conditions. 
The mean seeing during our observational campaign was approximately 2 arcseconds. As a comparison, the projected fibre diameter of the 2dF positioner is 2 arcseconds.

\subsection{Plate Configurations}\label{subsec:conf}

On each plate mounted on the 2dF positioner, out of the 392 fibers, about 350 are dedicated to observing science targets simultaneously across a 1.05-degree radius area in the sky. 
Around 30 fibers are reserved for observing blank sky regions, which is crucial for sky subtraction and calibration. To configure the fibers on the plates, we employ the \texttt{configure} software released by the observatory \citep{2006MNRAS.371.1537M}.

The sky fibers are essential for sky subtraction and assessment of weather conditions during the observations. For each field, we randomly selected around 100 locations devoid of any objects within a 5-arcsecond radius, which we visually verify. $\sim 30$ of those are typically used with sky fibers.
Meanwhile, guide stars are selected not only to guide the telescope, but also to determine the field plate rotation and set the relative position of the science fibers on the sky. We selected around 100 guide stars with $V$-band magnitudes within a range of either 13.0-13.5 or 13.5-14.0 (with a half-magnitude range chosen to allow even illumination across the different guide bundles) from the the Fourth U.S. Naval Observatory CCD Astrograph Catalog (UCAC4) \citep{ucac4} that are distributed uniformly across the FRB field. We also excluded stars with peculiar velocities larger than 0.015 arcsec per year to ensure the star positions have not drifted significantly since the publication of the UCAC4 catalog. Note that for each plate, only $\sim 6$ of those guide stars are utilized by the AAT system during the observation.

The science targets were selected using the criteria mentioned in Section \ref{subsec:selection}.
For the design of each plate using the \texttt{configure} software, we limit the input list of science targets to $\sim 800$ randomly-selected \wide-AAT science targets that have yet to be allocated, along with all the \narrow-AAT targets that have been allocated for less than 3 times.
We set this number of science targets to maximize fiber utilization while reducing the time required for the software to solve for an optimal fiber allocation configuration.
On subsequent plates in the same field, the input list to \texttt{configure} is ``topped up" with remaining targets in the source catalog to reach 800 science targets if available.
We repeated the procedure of allocating fibers and generating plate configurations until the usage of science fibers was lower than 50\% within a single plate, at which point we stop designing further plates for that field.

\subsection{Observing Log}\label{subsec:log}

Table \ref{logtab} shows the observing log of the 9 FRB fields we observed with AAT. For DR1, we observed 45 plates and 12948 targets in 9 FRB fields from 2020 through 2023. 
Most of the DR1 observations were achieved in 2020-2021, with a small fraction completed during 2022 and 2023 when most of the observations were for DR2.

Figure \ref{ob_frac} shows the completion rate of the AAT observations, for both the \wide-AAT and \narrow{}-AAT survey. The completion rate is defined by the fraction of observed targets divided by the total number of targets satisfying the selection criteria. We observed $\gtrsim70\%$ of the \wide-AAT targets for most fields. For \frbnine, the relatively low completion rate is due to unfavorable weather and equipment failure. 
Meanwhile, the low observing rate for the \frbthree{} \wide-AAT targets is due to the \textit{post-hoc} modification of the magnitude definition used for target selection after observations were carried out (see Section~\ref{subsec:selection}). 
However, because the \frbthree{} field is a comparatively low-redshift FRB ($\zfrb = 0.161$) covered by SDSS spectroscopy, the latter contributes enough foreground galaxies to reconstruct most of the foreground volume --- we therefore included this field in the DR1 analysis \citep{Khrykin2024flimflam} but will aim to improve the completeness in subsequent observations.
The observing rate of \narrow{}-AAT targets, on the other hand, is close to $80\%$ for all the fields except for \frbzero{}and \frbone. For these 2 fields, we only observed the brighter $\sim 40\%$ of the total targets but has already reached the dwarf galaxy depth as mentioned in Section \ref{subsec:structure}.

\begin{figure}[htbp]
\centering
\includegraphics[width=0.48\textwidth]{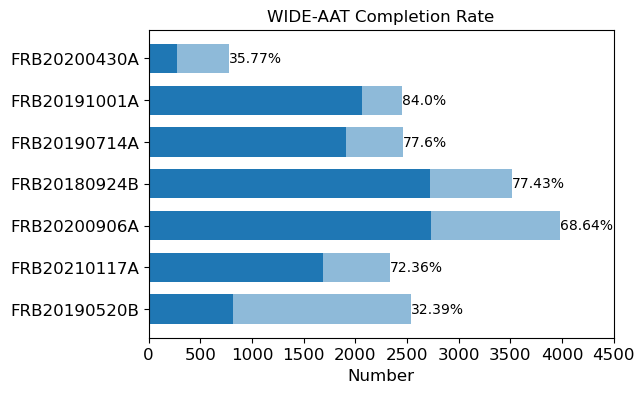}
\includegraphics[width=0.48\textwidth]{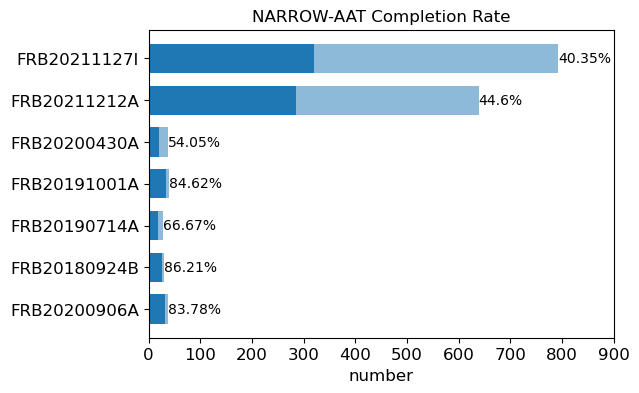}
\caption{(a) The observational completeness rate of the \wide-AAT survey. The bars with lighter color shows the amount of selected targets and the darker bars shows the amount of observed targets. The completion rate is showed on the right of each bar. Most of the FRB fields we have observed targets to $\gtrsim70\%$ completeness. (b) A similar plot for \narrow-AAT survey. The relatively low completion rate for \frbzero{}, \frbone{} and \frbthree{} are explained in Section \ref{subsec:selection}. For the rest fields, the completion rate is $\gtrsim80\%$. }
\label{ob_frac}
\end{figure}

\subsection{Data Reduction}\label{subsec:code}

To reduce the AAOmega data, we used a modified copy of \texttt{2dFDR} version 6.2 kindly provided by the Australian Dark Energy Survey (OzDES) group \citep{ozdes_dr1, ozdes_yr3, ozdes_dr2}. The processing involved utilizing the overlapped region to subtract bias and remove the cosmic ray by removing the affected pixels from subsequent analysis. This script maps the 1D spectra locations on the CCDs using fiber flats, subtracts the background light and extracts the flux using spline and Gaussian fits, with Gaussian widths determined by the flat fields. Wavelength calibration was performed using arc files, and sky emission lines were subtracted from target spectra using blank sky observations. The spectra of the same objects from multiple exposures were co-added with additional cosmic ray subtraction.

\begin{figure*}[htbp]
\centering
\includegraphics[width=0.95\textwidth]{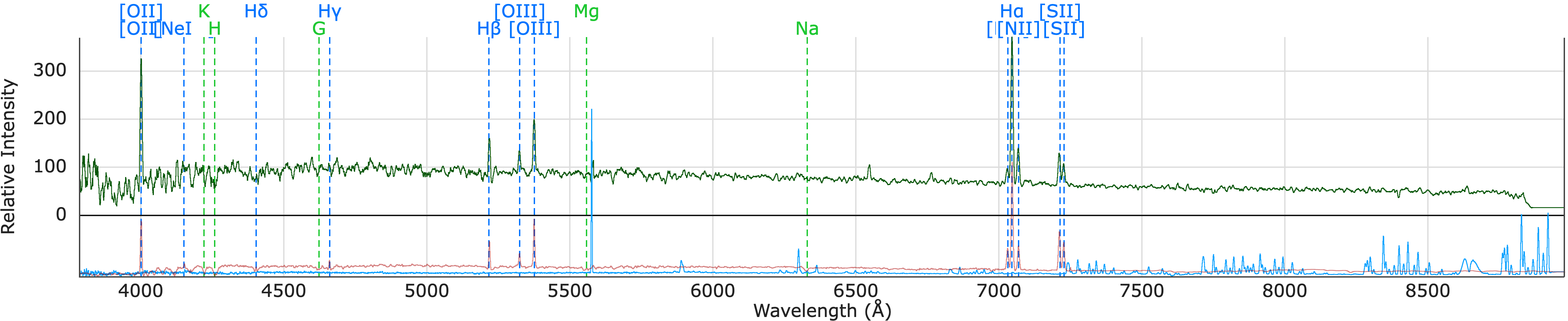}
\includegraphics[width=0.95\textwidth]{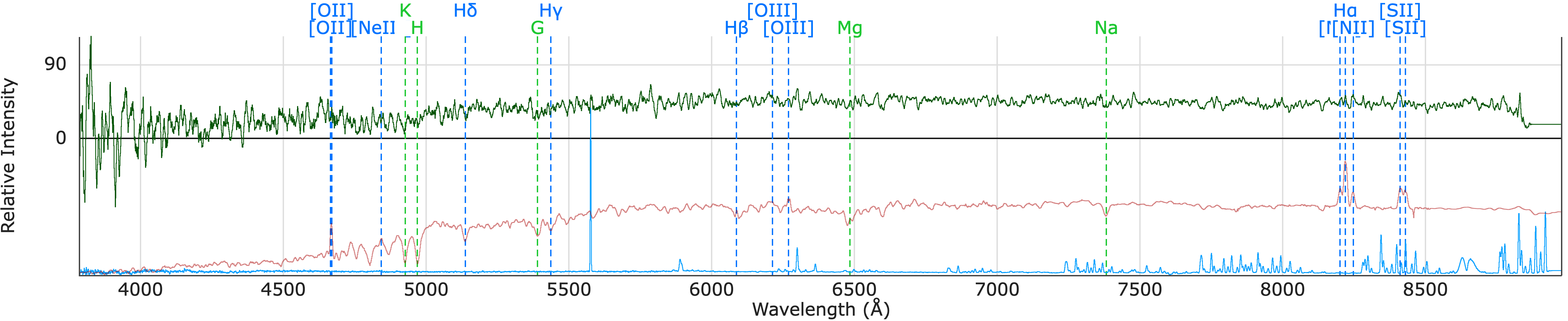}
\includegraphics[width=0.95\textwidth]{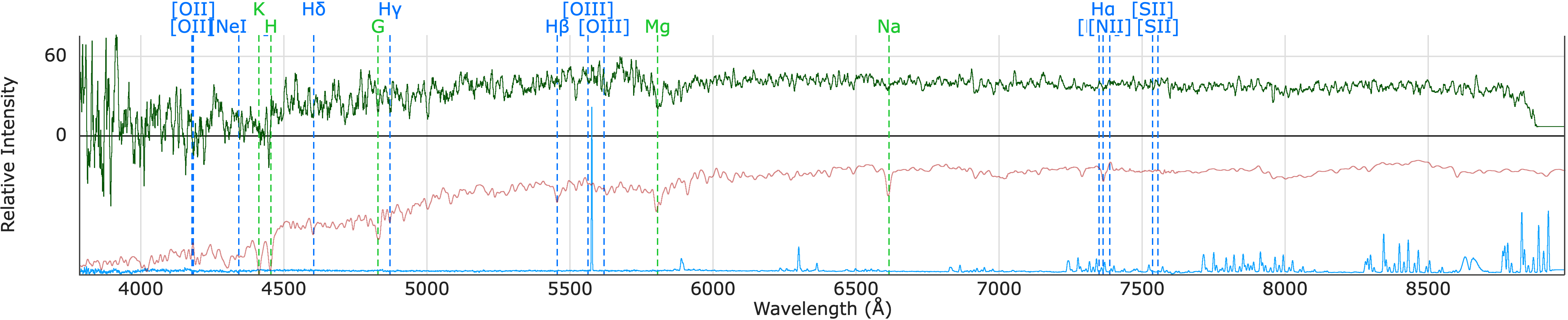}
\includegraphics[width=0.95\textwidth]{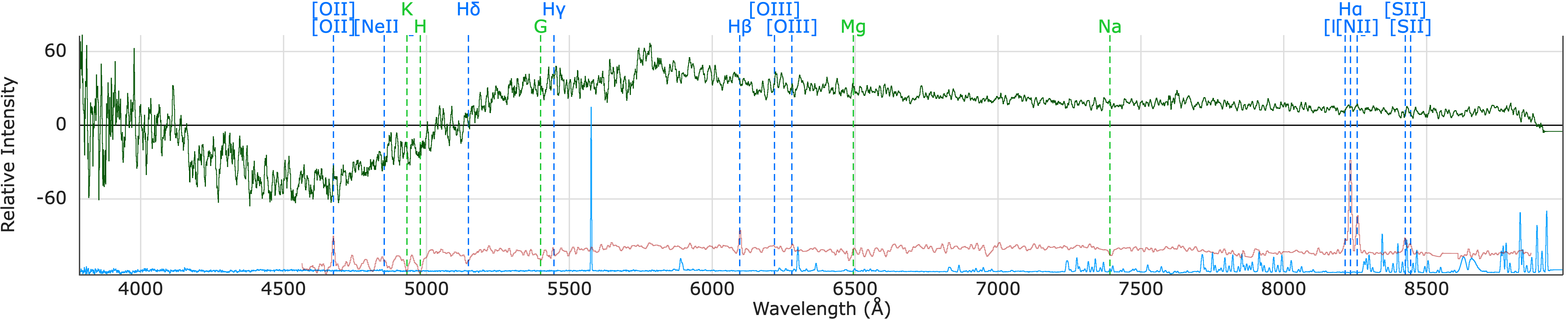}
\includegraphics[width=0.95\textwidth]{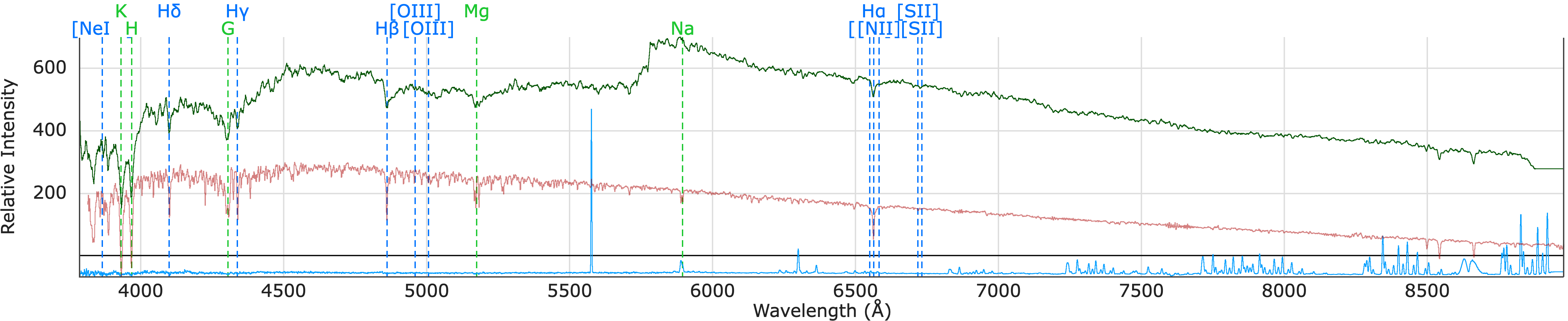}
\caption{Examples of typical QOP 4,3,2,1 and 6 spectra (from top to bottom). These spectra are selected from \frbeight{} field. In these plots, the green lines are the reduced spectra, the red lines are the corresponding templates given by \texttt{Marz}, and the blue lines are the blank sky spectra. The important emission lines and absorption lines are labeled in blue dashed lines. The top two spectra have been classified respectively as a late-type emission galaxy and an intermediate-type galaxy. Their characteristic hydrogen and metal line features allow for the determination of secure redshifts. The bottom spectrum is identified as a K star, a classification made evident by the clear metal absorption lines with no redshift.}
\label{qop}
\end{figure*}

\section{8-10 meter observations and data reduction}\label{sec:narrow_obs}
As described in Section \ref{subsec:selection}, we observed 7 FRB fields with 8-10 m class telescope spectroscopy. This section describes the observations and the data reduction.

\subsection{Observations}\label{subsec:slitmask}
We designed slitmasks for LRIS, DEIMOS, and GMOS-S to observe our deeper, narrow-field spectroscopic targets for 6 fields. Slitmasks are sheets of aluminum with slits milled at locations corresponding to targets on the sky. They enable simultaneous observations of multiple objects when placed in front of the spectrographs. Depending on the instrument field of view, each mask accommodates 25 (typical for LRIS and GMOS) to 50 (typical for DEIMOS) targets. Each slitmask was designed with 1 arcsec wide slits using the respective software for each instrument: \texttt{Autoslit} (by Judy Cohen, Patrick Shopbell, and Douglas Clowe) for LRIS, \texttt{DSimulator} (by Drew Phillips) for DEIMOS and \texttt{GMMPS} (by Bryan Miller et al.) for GMOS. In addition to our science targets, we also included $3''\times3''$ slits for alignment stars. Before science exposures, the slitmasks are aligned by centering pre-selected bright ($m_V=15-17$) stars within the square alignment slits.

\subsubsection{LRIS}
Keck LRIS, like AAOmega, has two channels for observations, i.e., a dichroic beam splitter that separates the incoming light into a blue and a red channel. We used the d560 dichroic, meaning channels separate at 5600 \AA. The blue channel was equipped with a 600/4000 grism and the red channel with a 600/7500 grating (central wavelength 7200 \AA). The blue channel has a wavelength coverage of 3330-5600\AA\ 
 (spectral FWHM $\sim4''$) and the red channel has a wavelength coverage of 5600-10000\AA{} (spectral FWHM $\sim4.7''$). Detectors in both chanels were binned $2\times2$. Each LRIS mask was exposed for 50 min to ensure emission lines could be detected for $r\sim23$ sources at $SNR/pixel>5$ and the continuum could be detected at $SNR/pixel\sim4$. For calibration, we obtained arc lamp exposures, flat field exposures, and bias frames for each mask. Using He, Ar, Xe, Kr, Cd, and Hg lamps, we obtained arc frames for the blue and red channels. We used the dome lamp for red-side flat fielding, while twilight flats were obtained for the blue channel. LRIS observations were carried out for the fields of \frbthree\ and \frbfive\ in 2020 before the survey design. Subsequently, these fields were re-observed with Keck DEIMOS to include missed targets in the previous LRIS run.

\subsubsection{DEIMOS}
The Keck DEIMOS instrument has a single channel with a wavelength coverage of 4500-10000\AA\ (spectral FWHM $\sim3.5''$). We used the 600ZD grating with a central wavelength of 6500 \AA{} as our dispersive element. We used $1\times1$ binning and set an exposure time of 50 min for each mask to ensure emission lines could be detected at $SNR/pixel>5$ for $r\sim23$ sources. Similar to LRIS, for calibration, we obtained bias, arc lamp exposures, flat field exposures and bias frames for each mask. We again used He, Ar, Xe, Kr and Cd and Hg lamps for arc frames (1s exposure) and the internal quartz lamp for flat fielding (4s exposure). All the narrow-field sightlines were observed with Keck DEIMOS with the exception of \frbfour\ on account of its southerly declination, thus making it inaccessible from Mauna Kea. 

\subsubsection{GMOS}
The Gemini South GMOS instrument also has a single channel with a wavelength coverage of 4000-11000\AA\ (spectral FWHM $\sim2.7$\,\AA). We used the B600 grating with two central wavelengths, 5500\,\AA\ and 6100\,\AA\ to cover the detector chip gaps. The detector was binned $2\times2$. For calibration, we obtained arc lamp exposures (Cu and Ar), flat field exposures, and bias frames. \frbfour\ is the only field that was observed with GMOS.
With each central wavelength, a mask was observed for $2\times 900$\,s to target $r<22.5$ sources thus, each object was observed for 3600\,s.

\subsubsection{MUSE}
The VLT/MUSE observation were conducted in its `wide-field' providing a field of view of $1\arcmin \times 1\arcmin$. We used the nominal wavelength range with adaptive optics, providing wavelength coverage of 
$4700-9300$\AA\footnote{with a $250$\AA\ gap centred at $\sim 5885$\AA.} (spectral FWHM $\sim2.5$\,\AA). For calibrations, we used files from the MUSE standard calibration plan which includes bias frames, flat field frames, and lamp exposures (Ne, Xe, and HgCd). The fields were observed for $8 \times 600$\,s to target $r\lesssim 24$ sources (total exposure time of 4800\,s).

\subsubsection{KCWI}
On UT 2019, September 30 and October 01, we obtained a 1 arcmin $\times$ 1 arcmin mosaic of 6 900 s exposures of the FRB20190608A field, centers on the host galaxy, with the KCWI on the Keck~II telescope. The data were obtained with the integral-field unit (IFU) in the ``Large" slicer position with the ``BL" grating, resulting in a field of view (FOV)
of $20'' \times 33''$ per pointing and a spectral resolution of R=900 (FWHM). Both observing nights were clear with seeing of FWHM$\sim0.9''$.

\subsection{Data Reduction}\label{subsec:reduc_narrow}
All the deeper data were reduced using the PypeIt reduction package \citep[v 1.2][]{pypeit}. 
PypeIt is a Python-based data reduction pipeline supporting several spectroscopic instruments, including LRIS, DEIMOS, and GMOS. All frames are first bias-subtracted, flat-fielded, and the cosmic rays are masked. Subsequently, PypeIt identifies the slit edges on the detector using the flats, identifies and subtracts sky emission features, identifies object traces within each slit, and extracts the 1D spectra. We used the optimal (``OPT'') extraction method for each object. The wavelength solution is estimated using arc lamp exposures. We used a combination of Mercury (Hg), Neon (Ne), Argon (Ar), Zinc (Zn), and Cadmium (Cd) arc lamps for the Keck LRIS and DEIMOS data. Copper-Argon (CuAr) lamps were used for Gemini GMOS. While PypeIt was able to successfully identify and label arc emission lines for Keck observations, we used the \texttt{pypeit\_identify} script to generate wavelength solutions manually for several objects in our GMOS multi-object spectroscopic data. This ensured better wavelength calibration (RMS error in line centers $\lesssim0.2$ pixels) than the default PypeIt reduction parameters. These solutions have been compiled and have been made available to users in subsequent versions of the software. PypeIt imposes a heliocentric correction, and the solutions are provided in vacuum. 

\begin{table*}[t] 
\centering
\begin{threeparttable}
\begin{tabularx}{0.82\textwidth}{c|ccccc}
\toprule
FRB & Observed Objects & Observed Targets & QOP 3 \& 4 targets & SSR & Completeness\\
\midrule
20211127I & 320 & 320 & 301 & 94.06\% & 38.59\%\\
20211212A & 285 & 285 & 276 & 96.84\% & 43.35\%\\
20200430A & 390 & 299 & 235 & 78.60\% & 15.35\% \\
20191001A & 2247 & 2096 & 1792 & 85.50\% & 73.59\% \\
20190714A & 2235 & 1927 & 1383 & 71.77\% & 59.43\%\\
20180924B & 2825 & 2746 & 2245 & 81.76\% & 65.37\%\\
20200906A & 2806 & 2763 & 2291 & 82.92\% & 57.83\%\\
20210117A & 1683 & 1406 & 1188 & 84.50\% & 55.65\%\\
20190520B & 1043 & 821 & 661 & 80.51\% & 26.71\%\\
\bottomrule
\end{tabularx}
\end{threeparttable}
\caption{The total number of observed objects, observed targets and quality flags of QOP 3/4 targets of each FRB field targeted with the AAT. We also indicate the spectroscopic success rate (SSR) and completeness of the observations.}
\label{suctab}
\end{table*}

PypeIt is capable of incorporating slit mask design information for multi-object spectroscopic reduction. This enables PypeIt to (1) predict the location of targeted objects within each slit on the detector and (2) assign object names and coordinates based on the input target tables provided to PypeIt. In the case of DEIMOS, this design information is stored in each raw FITS file and PyepIt is coded to read this information by default. LRIS and GMOS raw data do not have this information and thus must be pre-processed to incorporate it in a format recognizable to PypeIt. For LRIS, we used the TILSOTUA package to parse the AUTOSLIT mask design ASCII files and produce FITS tables that resemble those on DEIMOS raw frames. These were inserted manually into the raw LRIS flat frames. For GMOS, we directly used the GMMPS mask design FITS tables along with the WCS information encoded in each GMOS raw frame. To ensure that all objects, including the faintest sources, were identified and extracted by PypeIt, we set the `findobj $\rm snr_{thresh}$' parameter to 3.

Additionally, where targeted objects were still below this detection threshold, we forced object extraction at the location of the target from the mask designs. This produced wavelength calibrated 1d spectra for all targeted sources. Along with targeted sources, PypeIt identified several serendipitous sources. These are spectra of objects within the field of view which were not specifically targeted but were coincident with one of the mask slits. As we were only concerned with object redshifts, we did not flux calibrate the spectra.

As multiple exposures were obtained for each mask pointing, the individual 1d spectra were combined using the PypeIt collate1d script. The script pools all 1d spectra for each FRB field and collates objects by their assigned (R.A., Dec.) coordinates informed by the mask design. The collated spectra for each object are coadded automatically to produce a single FITS file for each object.

The VLT/MUSE data reduction was performed using the standard ESO MUSE Reduction pipeline (v2.8.6, \citealt{Weilbacher2020}) with standard parameters. The sky subtraction was improved using the ZAP code \citep{Soto2016}.

The KCWI data was similarly reduced with the KCWI Data Reduction Pipeline (KCWI DRP) v. 1.1.0 DEV 2019/05/23 with standard parameters. The pipeline produced datacubes with the wavelength solution provided in vacuum and in the heliocentric frame.

\section{Data Products} \label{sec:redshift}

After spectroscopic data reduction, we estimate redshifts of the targeted sources and subsequently perform calculations to aid our cosmic baryon analysis. We secured 10,330 reliable redshifts from the \wide-AAT survey, comprising 9820 galaxy redshifts and 510 stellar spectra. From the \narrow-AAT survey, we obtained in total 654 measurable redshifts, including 648 from galaxies and 6 stars.
Additionally, from the \narrow-8m and \ifu{} surveys, we obtained 839 redshifts, of which 813 are from galaxies and 26 are stars.

\subsection{Spectra to Redshifts}\label{subsec:spec}

The resulting 1D spectra were visually inspected by the FLIMFLAM group members using \texttt{Marz}\footnote{\href{https://samreay.github.io/Marz/\#/overview}{https://samreay.github.io/Marz/\#/overview}}\citep{Hinton_2016}. \texttt{Marz} is an open-source client-based Javascript web-application that is designed to estimate the redshift of the input spectra by computing the cross-correlation against its template spectral library. 
\texttt{Marz} assumes the input 1D spectra wavelengths is taken in air and shift them to the vacuum frame. It does not apply the heliocentric velocity correction to the spectra.

\begin{figure*}[htbp]
\centering
\includegraphics[width=0.9\textwidth]{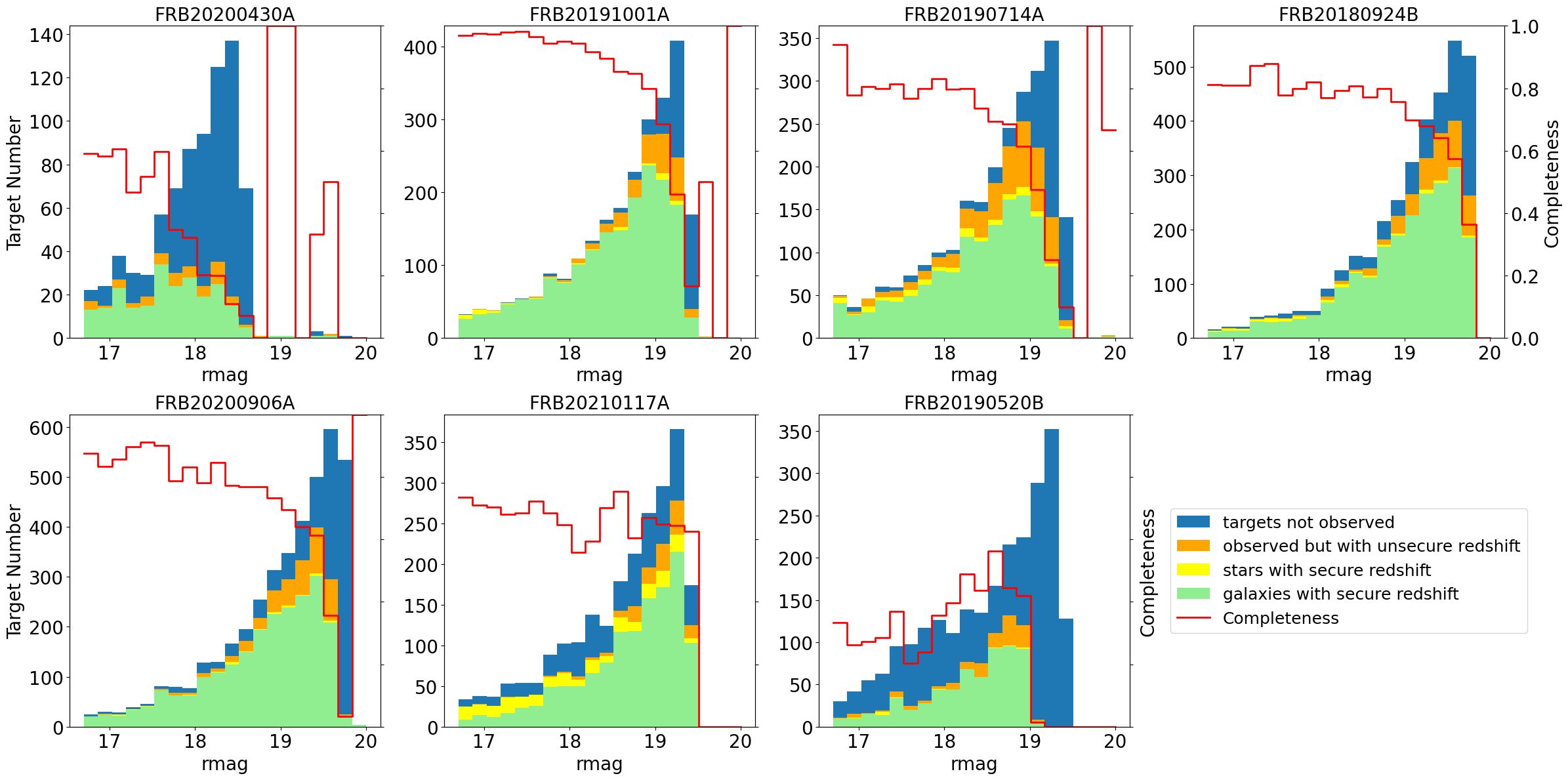}
\caption{The $r$-band distribution of galaxies with secure redshift (light green), stars with secure redshift (yellow), observed targets with unsecure redshift (orange) and targets not observed (blue) for the 7 \wide-AAT FRB fields. The completeness of each magnitude bin is also depicted with red step curves.}
\label{rmag_dist}
\end{figure*}

\texttt{Marz} also reports an automatic quality operator (AutoQOP) value representing the reliability of the resulting redshift for each matched spectrum. The AutoQOP is calculated from the two highest correlation values in terms of template and redshift combinations, and can take on values in [1,2,3,4,6]. AutoQOP 6 stands for a secure redshift prediction for a star, while AutoQOP from 1 to 4 represent galaxy redshifts from low to high confidence, respectively.

\begin{figure*}[htbp]
\centering
\includegraphics[width=0.9\textwidth]{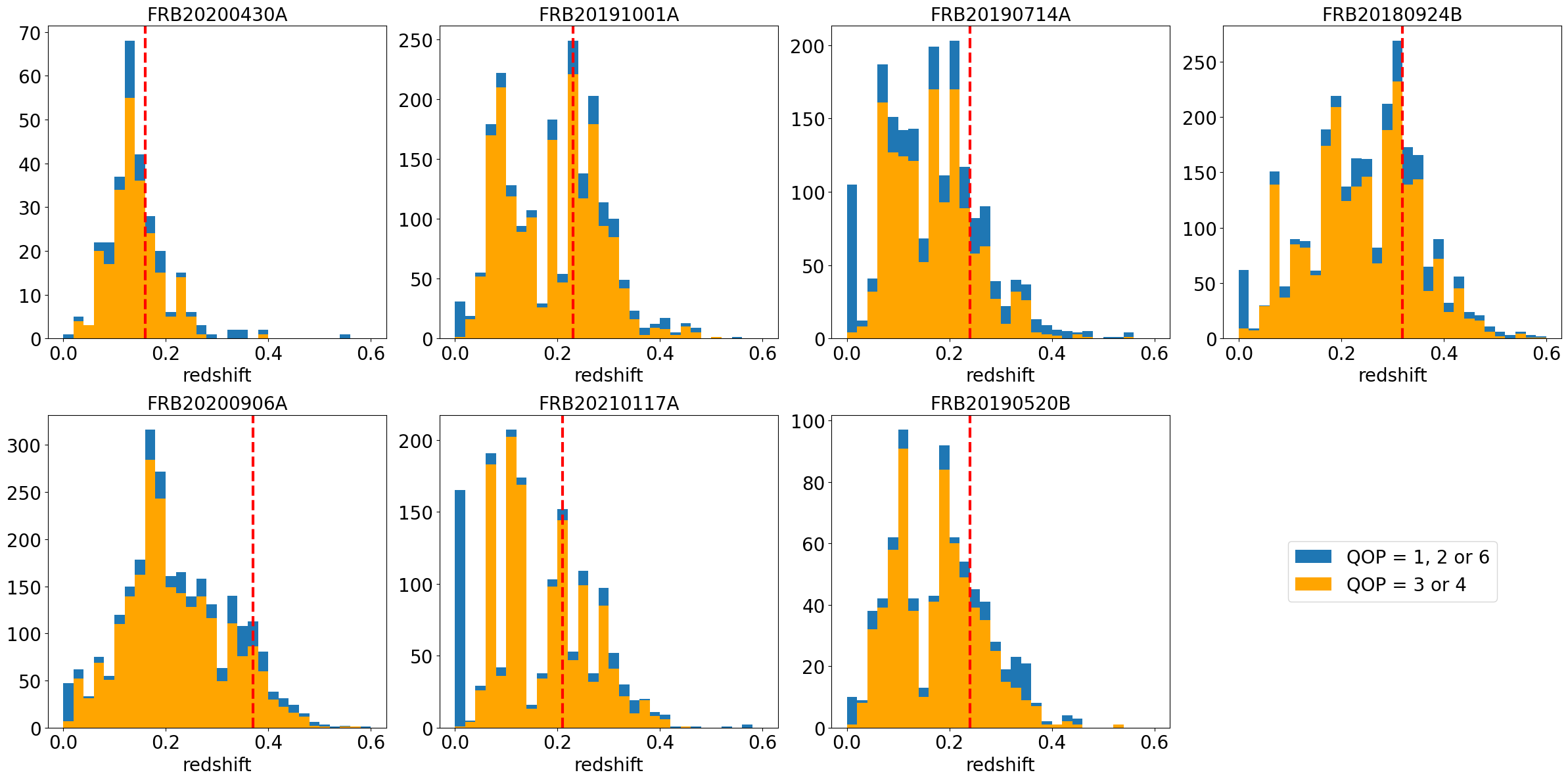}
\caption{The redshift distribution of the $7$ \wide-AAT FRB fields. The orange part shows the galaxies with reliable redshift. The red vertical dashed lines show the redshift of each corresponding FRB.}
\label{z_dist}
\end{figure*}

\texttt{Marz} was first used for redshift estimation in the OzDES survey \citep{ozdes_yr3} and has been shown to be fairly reliable \citep{Hinton_2016}. However for the FLIMFLAM spectra, we find that a small fraction of the spectra can be misidentified by \texttt{Marz}. For example, some spectra from stars are identified as galaxies, and some spectra with obvious emission or absorption features are flagged as AutoQOP 1 or 2, i.e. uncertain identification. 
To fully utilize our spectra and minimize misclassification, the FLIMFLAM group members visually inspected all the results of \texttt{Marz} and manually corrected the redshift as well as its QOP. We show typical QOP 4,3,2,1 and 6 spectra in Fig.\ref{qop}. The QOP 4 spectra typically have multiple strong spectroscopic features, like strong emission lines for late-type galaxies or strong metal absorption lines for early-type galaxies. The QOP 3 spectra have one strong spectroscopic feature or a few weak features. 
For QOP 3 and 4 spectra we have a high confidence on the classification and the resulting redshifts. 
The QOP 2 spectra only show 1 or 2 weak features, and QOP 1 spectra have no matching features with the templates, so the confidence of the given redshift is thus completely uncertain. Meanwhile, spectra with multiple strong features matching stellar templates and redshifts close to 0 are assigned QOP 6. For our cosmic baryon analysis in \citet{Khrykin2024flimflam}, only sources with QOP 3 and 4 spectra are used.

\begin{figure*}[htbp]
\centering
\includegraphics[width=0.9\textwidth]{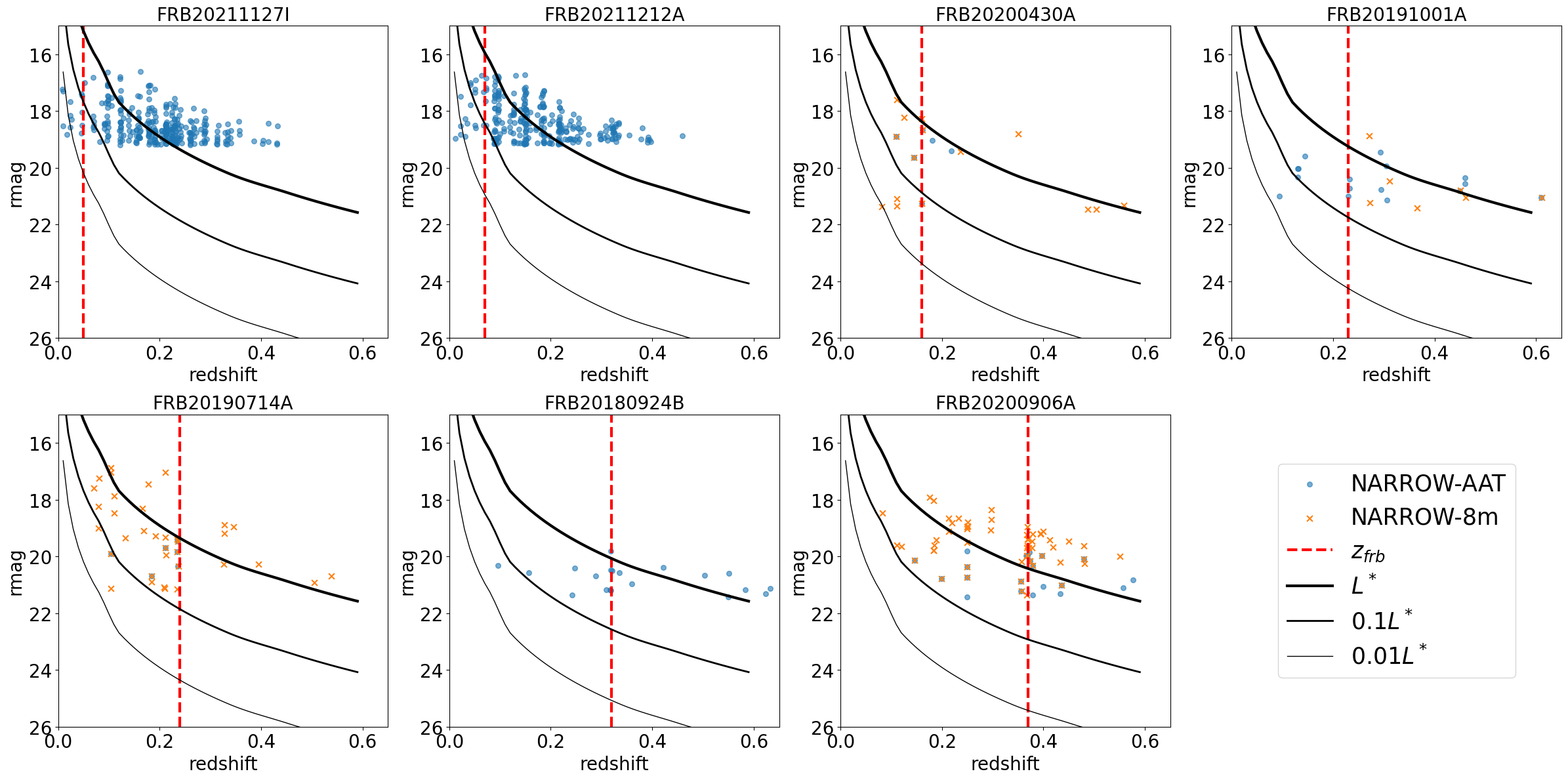}
\caption{The $r$-band magnitude as a function of redshift for the \narrow-AAT and \narrow-8m targets for the 7 FRB fields that were observed to study the cosmic baryon distribution. Here we only plot the targets that with QOP 3 or 4. The \narrow-AAT targets are labeled as blue dots and the \narrow-8m targets are denoted as orange crosses. For comparison, luminosity $L=L^{\star}$, $0.1L^{\star}$, and $ 0.01L^{\star}$ of the galaxy population at corresponding redshift is transferred into magnitude and labeled in black/gray/light-gray lines, respectively. The redshift of the corresponding FRB is indicated as a vertical red dashed line.}
\label{rmag_Lstar}
\end{figure*}

\subsection{Redshift Completeness and Distribution}\label{subsec:outcomes}

In this section, we first analyze the redshifts obtained from the \wide{}-AAT survey. With the spectra graded, and the redshifts derived, we list the number of observed objects, observed targets, QOP 3 and 4 targets, spectroscopic success rate and completeness in Table~\ref{suctab}. The spectroscopic success rate (SSR) is defined as the ratio between the observed targets with reliable redshift estimates (i.e., QOP 3 or QOP 4), $N_{\rm spec}^{\rm good}$, and the overall number of targets with acquired spectra, $N_{\rm spec}^{\rm all}$, i.e., ${\rm SSR} = N_{\rm spec}^{\rm good} / N_{\rm spec}^{\rm all}$. The completeness, on the other hand, is defined as the number of objects with secure redshift (QOP 3, 4 and 6) over the number of objects satisfying the selection criteria within the field, whether they were actually observed or not. 

For some fields, the number of observed targets differs from the number of observed objects. In other words, some of the objects, that were assigned fibers and observed, were subsequently removed from the formal FLIMFLAM target list. This is because we made \textit{post-hoc} changes to the magnitude definition used to define targets after observations were carried out, as described in Sec.\ref{subsec:selection}. \frbthree{} is the only field that has been significantly influenced by the change of magnitude definition, and we will observe 1-2 more plates on this field in future campaigns.

The spectroscopic success rate across all fields is notably high (approximately $80\%$). The overall completeness is around $50\%$ on average but varies significantly, ranging from $20\%$ to $70\%$ among different fields. This drop is due to multiple reasons. For \frbzero{} and \frbone{} we only observe the brighter half of the selected targets, however we show in Fig.\ref{rmag_Lstar} that the galaxies we observed is sufficient for cosmic baryon analysis. 
The excess DM analysis of \frbeight{} and \frbnine{} by \cite{Simha2023} and \cite{Lee_2023} 
respectively only required us to identify large galaxy groups or clusters in the foreground and hence do not require high completeness. 

In Fig.\ref{rmag_dist} we plot the dereddened $r$-band magnitude distribution of galaxies with secure redshifts, confirmed stars, observed targets with unsecure redshift and targets not observed. From the plot, it is evident that the observation depth and completeness of \frbzero, \frbone, \frbthree{} and \frbnine{} do not reach their respective nominal $r$-band cuts due to bad weather conditions, the lack of observation time, and the change in magnitude definition for target selection. The success rate is high ($\gtrsim90\%$) for $r<18.6$ targets and decrease with increasing magnitude. 
Nevertheless, the success rates of $r>19.2$ targets is generally above $80\%$.

The redshift distributions of the nine fields from the \wide-AAT observations are showed in Fig.\ref{z_dist}. The redshift of each FRB is labeled as vertical red dashed line in the plots. We generally achieve a high success rate for targets with redshift smaller than 0.4. Also the bumps at redshift smaller than $z_{\rm FRB}$ of each histogram represent the existence of intervening large-scale structure and galaxy halos, which could make a large contribution to the total DM. The detailed analysis is presented in \cite{Khrykin2024flimflam}.

As for the \narrow{} and the \ifu{} observations, our nominal goal is to reach galaxy luminosity corresponding to $0.1L^{\star}$ in the foreground. In Figure~\ref{rmag_Lstar}, we show the $r$-band magnitude as a function of redshift for \narrow-AAT and \narrow-8m targets. The curves representing $L=L^{\star}$, $0.1L^{\star}$, and $ 0.01L^{\star}$, as a function of redshift, are also shown for comparison. 
The $L^{\star}$ of corresponding redshift is calculated from luminosity functions of corresponding redshift and photometric bands \citep{Brown_2001, Willmer_2006, refId0}; we
refer to \citet{Heintz_2020} for the method. From the figure, we show that our \narrow-AAT and \narrow-8m spectroscopy have generally achieved depths of $0.1L^{\star}$ for targets in the foreground of our FRBs, and for the two low-redshift fields \frbzero{ }and \frbone, we have even reached $\sim0.01L^{\star}$.

To provide a visual impression of the lightcone fields we have observed, we plot the projected scatter plot of FLIMFLAM galaxies in Cartesian coordinate of all the 10 DR1 FRB fields in Fig.\ref{lightcones}.

\begin{figure*}[htbp]
\centering
\includegraphics[width=0.9\textwidth]{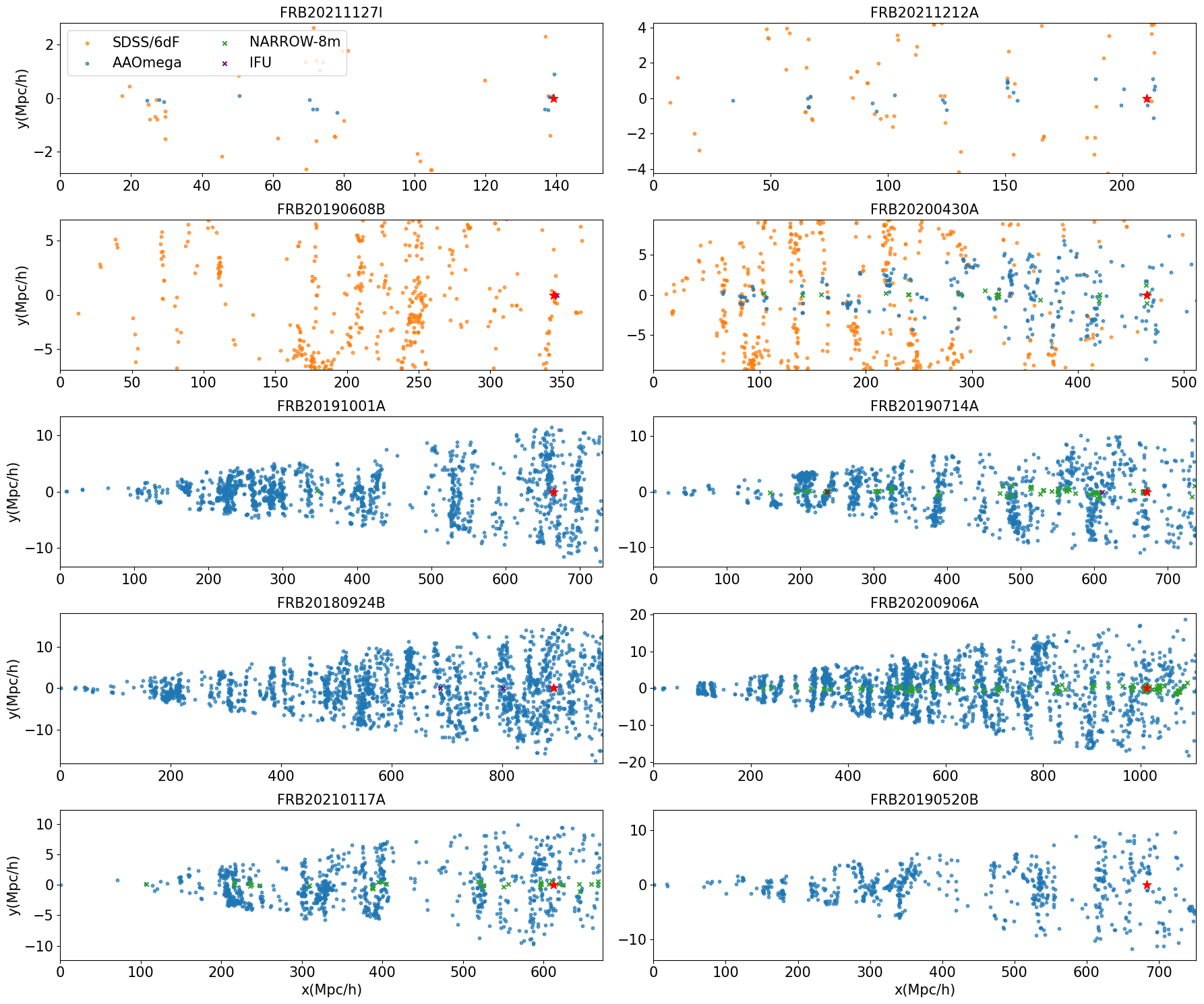}
\caption{Projected scatter plot of galaxies in the FRB foreground fields in Cartesian coordinate. The $x$-axis is the line-of-sight direction of each FRB sightline and the $y$-axis is the transverse direction of the sightline in the Right Ascension dimension. The red star represents the position of each FRB, and the foreground AAOmega galaxies (including both \wide-AAT and \narrow-AAT galaxies), SDSS/6dF galaxies, galaxies obtained from \narrow-8m observations and galaxies from \ifu{} observations are labeled as blue dots, orange dots, green crosses and purple crosses respectively. For \frbzero, \frbone, \frbtwo{} and \frbthree{} fields which include the \wide-SDSS and the \wide-6dF galaxies, we only plot those galaxies with Declination (i.e. the dimension going into the page) within 1.05 deg from the FRBs. }
\label{lightcones}
\end{figure*}

\subsection{Selection Functions}
\label{sec:selfunc}

\begin{figure}
    \centering
    \includegraphics[width=0.9\columnwidth]{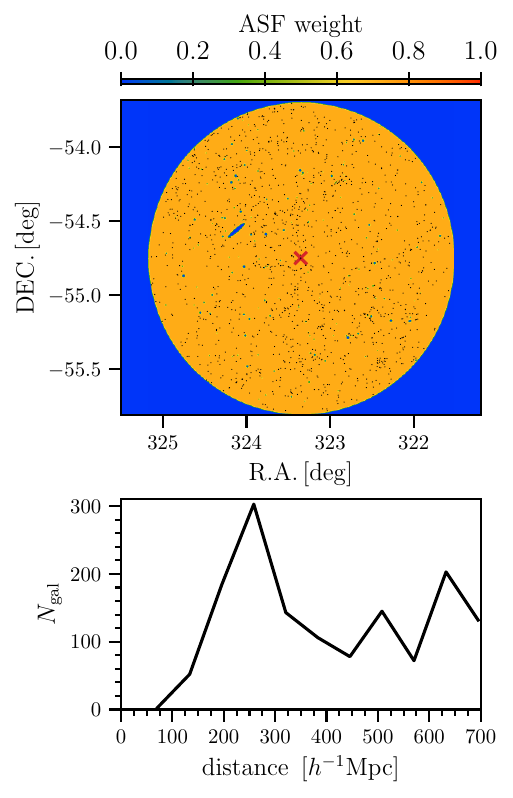}
    \caption{{\it Top}: the angular selection function (ASF) estimated in the field of FRB~20191001A. The black dots show the positions of the observed galaxies, while the position of the FRB is marked by the red cross.
    The prominent ellipse at (R.A.,Dec.)=(323.2469,-54.7817) deg is the nearby galaxy NGC7090.  {\it Bottom}: the corresponding radial selection function in the same observed field. The RSF is computed in bins of $10~h^{-1}$~Mpc.}
    \label{fig:selfuncs}
\end{figure}

Having acquired the final galaxy samples in each FRB field, we estimate the angular and radial selection functions (ASF and RSF, respectively) from these data. This information, on the observational completeness, is utilized by the matter density reconstruction algorithm \texttt{ARGO} \citep{Ata_2015, Ata_2017, Ata_2021}, adopted in the FLIMFLAM survey, to, e.g., resolve whether gaps in the observed galaxy distribution are due to cosmic voids or due to the lack of observations \citep[see][]{Lee_2022, Khrykin2024flimflam}.

To estimate the ASF in the circular field with $1.0~{\rm deg}$ radius around the on-sky position of a given FRB, we adopt the algorithm described in \cite{Ata_2021}, which we briefly outline here. The process of ASF calculation requires information from two steps of the target selection, from building an initial photometric sample, to the final high-quality spectroscopic sample. 
After reducing the observational data (see Section~\ref{sec:obs}), first, we estimate the so-called {\it target sampling rate} (TSR). It is defined as the ratio between the number of targets for which the spectra were acquired (regardless of the actual quality, in contrast to the completeness discussed in Section~\ref{subsec:outcomes}), $N_{\rm spec}^{\rm all}$, and the number of photometric targets in the source catalog, $N_{\rm phot}^{\rm src}$, ${\rm TSR} = N_{\rm spec}^{\rm all} / N_{\rm phot}^{\rm src}$. Then, we calculate the SSR as defined in Section~\ref{subsec:outcomes}. The TSR and SSR are both evaluated uniformly over the 3.1 deg$^2$ AAOmega footprint. 

In addition, to account for possible gaps in the field caused by saturation or diffraction spikes from bright stars, we construct the angular selection masks (ASM) by utilizing the imaging information from NOIRLab or Pan-STARRS surveys. Each mask covers an area of $1.0~{\rm deg}^2$ around the position of a given FRB and is set to unity by default. Next, using the 2MASS All-Sky Point Source, 2MASS All-Sky Extend Source and GAIA DR3 catalogues, we identify bright stars ($g < 15$) and galaxies ($J{\rm -band} < 15$) in the field, and draw a circular region around each one. The radius of a drawn region is proportional to the observed magnitude of an object. For exceptionally bright galaxies that appear elliptical in the image and are not adequately covered by circular masks, we use elliptical masks instead. Inside these regions, the mask value is set to be zero.

The resulting angular completeness mask in a given field is then given by 

\begin{equation}
    {\rm ASF} = {\rm TSR} \times {\rm SSR} \times {\rm ASM}.
\end{equation}

The top panel of Figure~\ref{fig:selfuncs} illustrates the final ASF, estimated in the field of FRB~20191001A, whose location is marked by the red cross. The black dots indicate the on-sky position of the AAT observed galaxies with high-quality redshifts. The overall ASF weight in this field is estimated to be $w \simeq 0.7$ \citep{Ata_2015,Ata_2017,Ata_2021}.

Finally, we estimate the RSF, which represents the number counts of galaxies as a function of distance (or redshift $z$) from the observer. To calculate the RSF, we consider only galaxies with high-quality redshift measurements and convert their redshifts into the corresponding comoving distances. The line-of-sight distribution of galaxies is estimated in bins of $0.1$~Mpc for each field FRB field in our sample. An example RSF in the field of FRB~20191001A is illustrated in the bottom panel of Figure~\ref{fig:selfuncs}. It is apparent that the distribution peaks around $z\simeq 0.08$. 

Both the ASF and RSF are combined to produce the so-called response function that is later projected into the \texttt{ARGO} density reconstruction volume Cartesian coordinate grid \citep[see][]{Lee_2022, Khrykin2024flimflam}.

\section{Conclusion} \label{sec:conclusion}

In this work, we present the first data release collected with the 2dF-AAOmega fiber spectrograph on the Anglo-Australian Telescope as part of the FLIMFLAM spectroscopic survey of galaxies, aimed at mapping out the large-scale cosmic structures in the foreground of localized FRBs. 
The data will be made publicly available on Zenodo after the review process.

The FLIMFLAM DR1 includes $10$ FRB fields (see Table~\ref{FRB_info}), where we have conducted spectroscopic observation both in the \wide- and \narrow-fields. 
The former covers the area of $\approx 26~{\rm deg}^2$ around the on-sky position of the corresponding FRB, and delivers the samples of galaxies that serve as tracers of the large scale matter density field (see \cite{Lee_2022, Khrykin2024flimflam}). The latter, on the other hand, is designed to probe the possibly-intervening galaxy halos (up to $r_{\rm mag} \leq 23$) over $\approx 137~{\rm arcmin}^2$ area from the FRB sightlines. 

Our DR1 observational campaign has observed $13737$ objects and yields $11281$ reliable galaxy redshifts and $542$ secure stellar identifications in total across all FRB fields in our sample, both in the \wide-, \narrow- and \ifu-fields. We report a $\approx 80\%$ spectroscopic success rate achieved in the \wide-field observations for most of the targeted FRB fields, and $>90\%$ success rate in the \narrow-field observations, respectively. The overall target completeness is $\sim50\%$ on average and varies between $20\%$ to $70\%$ among different fields. 
Two of the fields (\frbeight{} and \frbnine{}) are not included in the DR1 cosmic baryon analysis, but were instead analyzed as ``excess-DM" fields \citep{Simha2023,Lee_2023} with lower demands on their completeness.
Nevertheless, improving the completeness in the other five fields can reduce the uncertainty in the constraints of IGM and intervening halo DM \citep[see][for details]{Khrykin2024flimflam}. Therefore, should spare observation time become available in the future, we plan to observe an additional 1-2 plates for these fields.
The relatively low completeness of \frbzero, \frbone, and \frbthree{} do not significantly affect our cosmic baryon distribution constraint: the first two are very low-redshift ($\zfrb<0.1$) FRBs where even relatively shallow AAT observations reach dwarf-galaxy luminosities ($L<0.1\,L^*$) in the \narrow{} sample, while the incomplete \wide-AAT data on \frbthree{} is supplemented by SDSS spectroscopy. 
For the other fields, the \narrow{} spectroscopy have also generally reached depths corresponding to dwarf galaxies ($L \sim 0.1L^{\star}$) at the foreground of each FRB. 
This depth and completeness allows us to nearly complete census of the DM contributions on our fields.

Part of the DR1 data has been analyzed in \citet{Lee_2023} and \citet{Simha2023} before the analysis of the cosmic baryon distribution \citep{Khrykin2024flimflam} published this year. \citet{Simha2023} estimated the DM contributed by intervening halos for \frbthree, \frbseven, \frbfive{} and \frbeight{} by modeling the gas distribution around the foreground halos. 
They find a large halo DM that contributes 2/3 of the excess DM for \frbfive{} and a smaller but non-negligible foreground halo DM contributions for \frbseven. The other two FRBs do not exhibit significant foreground contributions, suggesting that they have above-average host contributions.
\citet{Lee_2023}, meanwhile, analyzed the foreground contributions of \frbnine{}, which has a very large DM (${\rm DM}\approx 1200\,\dmunits$) that was previously attributed to a large host contribution ($\dmhost\approx900\,\dmunits$; \citealt{Niu2022,Koch_Ocker_2022}). 
They end up finding 2 separate $\mhalo>10^{14}M_{\odot}$ galaxy clusters, which revise the host contribution downwards to $280-430\dmunits$ (no longer the largest-known \dmhost). These 2 studies illustrates the importance of understanding foreground data when analyzing the source of DM excess and using FRBs to solve cosmological problems.

In the coming years, the initiation of the CRAFT Coherent upgrade project (CRACO) on the ASKAP, as well as other survey such as DSA-2000, is expected to significantly enhance detection rates of localizable FRBs, increasing from one FRB per month to one FRB per day. Consequently, we anticipate detecting over 100 FRBs per year, which will greatly improve the analysis of cosmic baryon distribution. Furthermore, with the impending public availability of the Dark Energy Spectroscopic Instrument (DESI) and the Subaru Prime Focus Spectrograph (PFS), we expect a significant reduction in required observation time — from over 100 hours per field to just a few hours. Additionally, these advancements will enable us to extend our target FRBs to $z>1$. 

Pan-STARRS DR1 data used in this paper can be accessed via MAST: \href{http://dx.doi.org/10.17909/55e7-5x63}{10.17909/55e7-5x63}. GAIA DR3 data are available through IPAC: \href{http://dx.doi.org/10.26131/IRSA544}{10.26131/IRSA544}. 2MASS All-Sky Point Source data can be found at IPAC: \href{http://dx.doi.org/10.26131/IRSA2}{10.26131/IRSA2}. 2MASS All-Sky Extended Source data are also accessible via IPAC: \href{http://dx.doi.org/10.26131/IRSA97}{10.26131/IRSA97}.


\begin{acknowledgments}
We would like to express our gratitude to Chris Lidman for his assistance in setting up the environment for the reduction code and testing our data. YH acknowledges the supports from the University of Tokyo Global Science Graduate Course program. I.S.K. and N.T. would like to acknowledge the support received by the Joint Committee ESO-Government of Chile grant ORP 40/2022. MG is supported by the Australian Government through the Australian Research Council’s Discovery Projects funding scheme (DP210102103). ATD acknowledges support through Australian Research Council Discovery Project DP220102305. LM is supported by an Australian Government Research Training Program (RTP) Scholarship. RMS acknowledges support through Australian Research Council Future Fellowship FT190100155.
We acknowledge generous financial support from Kavli IPMU that made FLIMFLAM possible. Kavli IPMU is supported by World Premier International Research Center Initiative (WPI), MEXT, Japan. Based on data acquired at the Anglo-Australian Telescope, under programs A/2020B/04, A/2021A/13, and O/2021A/3001, we acknowledge the traditional custodians of the land on which the AAT stands, the Gamilaraay people, and pay our respects to elders past and present.
Some of the data presented herein were obtained at the W.M. Keck Observatory, which is
operated as a scientific partnership among the California Institute of Technology, the University of California and the National Aeronautics and Space Administration (NASA). The Observatory was made possible by the generous financial support of the W.M. Keck Foundation. The authors also wish to recognize and acknowledge the very significant cultural role and reverence that the summit of Maunakea has always had within the indigenous Hawai’ian community. We are most fortunate to have the opportunity to conduct observations from this mountain. 
Partially based on observations collected at the European Southern Observatory under ESO programmes: 
2102.A-5005(A), 0104.A-0411(A), 
105.20HG.001, 
110.241Y.001, and 
110.241Y.002. 
Partially based on observations obtained at the international Gemini Observatory (PID: GS-2022B-Q-137), a program of NSF NOIRLab, which is managed by the Association of Universities for Research in Astronomy (AURA) under a cooperative agreement with the U.S. National Science Foundation on behalf of the Gemini Observatory partnership: the U.S. National Science Foundation (United States), National Research Council (Canada), Agencia Nacional de Investigaci\'{o}n y Desarrollo (Chile), Ministerio de Ciencia, Tecnolog\'{i}a e Innovaci\'{o}n (Argentina), Minist\'{e}rio da Ci\^{e}ncia, Tecnologia, Inova\c{c}\~{o}es e Comunica\c{c}\~{o}es (Brazil), and Korea Astronomy and Space Science Institute (Republic of Korea).

\end{acknowledgments}




\bibliography{sample631}{}
\bibliographystyle{aasjournal}



\end{document}
